%% file: SourceCoding.tex
\documentclass[journal]{IEEEtran}
\usepackage{cite}
\usepackage{amsmath,amssymb,amsfonts,amsthm}
\usepackage{graphicx,color,xcolor}
\usepackage{tikz}
\usetikzlibrary{calc}
\definecolor{darkgreen}{RGB}{0,100,0}
\usepackage{textcomp}
\usepackage{etoolbox}
\usepackage{float} 
\usepackage{subcaption}
\usepackage[dvipsnames]{xcolor} 
\definecolor{ZoneA}{RGB}{31,119,180}  
\definecolor{ZoneB}{RGB}{255,127,14}  
\definecolor{ZoneC}{RGB}{148,103,189} 
\usepackage{pgfplots}
\pgfplotsset{compat=1.18}
\usepackage{pgfplots}
\usepgfplotslibrary{groupplots,fillbetween}

\usepackage{algorithm,algorithmicx}
\usepackage{algpseudocode}
\makeatletter
\AtBeginEnvironment{algorithmic}{%
  \footnotesize                       
  \setlength{\itemsep}{0pt}%
  \setlength{\parskip}{0pt}%
  \setlength{\topsep}{1pt}%
  \setlength{\partopsep}{0pt}%
  \setlength{\ALG@thistlm}{\algorithmicindent}%
}
\algrenewcommand\alglinenumber[1]{\scriptsize\makebox[0.9em][r]{#1}}
\makeatother
\usepackage{comment} 

\usepackage{pgf}
\usepackage{tikz}
\usetikzlibrary{arrows.meta,bending,positioning}
\usetikzlibrary{shapes,shadows,arrows,positioning}
\usetikzlibrary{graphs}

\tikzset{
    node/.style={circle, draw, thick, minimum size=3mm, inner sep=0pt, fill=white},
    baseedge/.style={line width=0.4pt, draw=gray!50},
    p1/.style={line width=2.2pt, draw=blue},
    p2/.style={line width=2.2pt, draw=darkgreen},
    p3/.style={line width=2.2pt, draw=orange!90!black},
    ovl/.style={line width=3.0pt, draw=black, densely dotted},
    every label/.style={font=\footnotesize}
}

\begin{document}
\title{Source-Coded Online Algorithm for Multicast Subgraph Construction}

\author{Tomás Lestayo Martínez\IEEEauthorrefmark{1}, Manuel Fernández 
Veiga\IEEEauthorrefmark{2} \IEEEmembership{(Senior Member, IEEE)}}
\maketitle


\begin{abstract}
Multicast remains a fundamental mechanism for scalable content distribution, yet 
existing approaches face critical limitations. Traditional multicast trees suffer 
from path redundancy and inefficient utilization of network resources, while network
coding, although capacity-achieving, incurs significant computational overhead and
deployment challenges. In this paper, we introduce a source-coded multicast framework 
that exploits maximum-flow decomposition to construct multiple disjoint or partially
overlapping paths from the source to all receivers. Our scheme incorporates a novel 
path redirection mechanism: when multiple overlaps occur between receiver flows,
downstream paths are realigned at the first intersection, ensuring loop-free delivery
while maximizing overall throughput. We develop algorithms for path construction, 
overlap detection, and iterative refinement of multicast subgraphs, and analyze their 
computational complexity. Through extensive evaluation on synthetic and real network 
topologies, we demonstrate that the proposed method consistently approaches the 
throughput of network coding with substantially lower encoding and decoding 
complexity, while significantly outperforming multicast tree constructions in terms 
of fairness, robustness to link failures, and delivery efficiency. These results 
position source-coded multicast as a practical and scalable solution for 
next-generation networks requiring high-throughput and adaptive group communication.
\end{abstract}


\section{Introduction}
\label{sec:intro}

\IEEEPARstart{M}{ulticast} communication has long been recognized as a cornerstone 
of scalable content delivery in modern communication networks. Applications such as 
IPTV, real-time video streaming, distributed computing, and large-scale data 
dissemination rely on the ability to efficiently deliver identical content from a
single source to multiple receivers. Traditional multicast mechanisms typically 
construct rooted trees that span the source, ensuring minimal duplication of traffic
across the network. Although effective in simple scenarios, these tree-based approaches
are inherently limited by their rigidity: they often fail to fully exploit path 
diversity, are highly sensitive to link bottlenecks, and provide limited resilience 
to failures~\cite{moy1994multicast,widmer2001extending,chu2002case}.

To overcome these limitations, a natural formulation is to compute multiple 
edge-disjoint or partially overlapping paths between the source and each receiver, 
thereby exploiting network diversity. This reduces to the maximum flow problem between 
a single source and a single sink. The classical framework is the Ford–Fulkerson
method~\cite{Ford1956}, later refined by Dinic’s algorithm~\cite{Dinic1970} and 
the Edmonds–Karp algorithm~\cite{EdmondsKarp1972}, which provide polynomial-time
solutions. In this work, we adopt Edmonds–Karp as our baseline: its use of 
breadth-first searches to identify augmenting paths aligns well with our online 
design, and it performs efficiently in the unit-capacity networks considered here.

Network coding has emerged as a theoretically optimal solution, achieving the 
multicast capacity by allowing intermediate nodes to forward linear combinations of 
data symbols in transit toward the next hop~\cite{ahlswede2000network,Li2003,Koetter2003}.
However, despite its elegance, the practical deployment of network coding has been 
hampered by challenges such as computational overhead, synchronization requirements, 
and integration with forwarding infrastructures~\cite{chou2003practical,ho2006random}.
These limitations motivate the search for alternatives that retain part of the 
performance gains of coding while preserving the simplicity of path-based forwarding.

In this work, we propose an \emph{online source-coded multicast framework} that
incrementally builds receiver flows through color-constrained breadth-first searches.
Instead of recomputing maximum flows globally with Edmonds–Karp for each new receiver, 
our algorithm processes receivers sequentially, enforcing \emph{color constraints} 
that control how paths may overlap. Each color represents a flow substream assigned to 
a receiver path; a receiver never reuses the same color across its paths, and any 
overlap with existing flows is limited to a single color. When a newly added path
intersects multiple existing flows, it is redirected at the first overlap point, 
ensuring loop-free construction and feasibility. Crucially, these constraints align 
with the conditions identified in the 
works~\cite{lestayo2001adaptive,Lestayo2023,lestayo2024source}, which show that 
such overlap patterns remain decodable under source-only coding, without requiring 
in-network operations. This design eliminates global recomputation, adapts naturally 
to dynamic receiver arrivals, and scales efficiently with network size.

For comparison purposes, we also construct network coding–optimal multicast subgraphs 
by applying Edmonds–Karp independently for each receiver and merging the resulting 
flows. This provides a benchmark that isolates the structural advantage of coding 
from the computational overhead of symbol mixing, and allows us to contextualize 
the performance of our online framework relative to both classical trees and 
coding-enabled multicast.

The main contributions of this paper are summarized as follows:
\begin{enumerate}
\item We introduce an online source-coded multicast framework that incrementally 
constructs edge-disjoint or color-constrained overlapping paths from the source to 
receivers.

\item We formalize a color-constrained breadth-first search that enforces overlap 
consistency, ensuring feasibility and compatibility with source coding at the sender.

\item We design algorithms for online path integration and redirection at intersections, 
enabling multicast subgraphs to be built without global recomputation.

\item We establish a coding-optimal benchmark by running Edmonds–Karp per receiver 
and merging flows, providing a natural baseline to assess coding gains.

\item We evaluate the framework on synthetic and real network topologies, showing that 
it achieves near-optimal throughput compared to network coding with substantially lower
complexity, and that it outperforms multicast trees in terms of fairness, resilience, 
and link utilization.
\end{enumerate}

Extensive evaluations confirm that the proposed online framework offers a scalable 
and practical solution for content distribution. Our results highlight the potential 
of combining source-based flow decomposition with lightweight online integration 
rules, yielding a favorable trade-off between efficiency, robustness, and deployability.

The remainder of this paper is organized as follows. Section~\ref{sec:related-work} 
reviews related work on multicast tree construction, network coding, and source-coded
multicast. Section~\ref{sec:model} introduces the system model and problem formulation.
Section~\ref{sec:framework} details the proposed online framework and algorithm. 
Section~\ref{sec:algorithm} analyzes complexity and performance properties. 
Section~\ref{sec:results} presents evaluation results and benchmarks. Finally, 
Section~\ref{sec:conclusion} concludes the paper.

\subsection{Summary of notations}
For clarity, we summarize the notation used throughout this paper. $G = (V,E)$ denotes a directed network graph with vertex set $V$ and edge set $E$. The capacity of edge $(u, v) \in E$ is symbolized by $c(u,v)$. The letter $s$ will generally be used to refer to the source node, and $\mathcal{R} = \{r_1, \ldots, r_m\}$ for a  set of receivers. The maximum flow from $s$ to receiver $r_i$ is written as $f(s,r_i)$. $\mathcal{P}_i$ is the set of paths carrying flow to receiver $r_i$, and $\mathcal{G}_{\text{multicast}}$ is the multicast subgraph formed by the union of $\mathcal{P}_i$.

\section{Related Work}
\label{sec:related-work}

The problem of efficiently delivering identical content from one source to multiple
receivers has been extensively studied in the networking literature. Early solutions
focused on multicast tree construction, with approaches such as shortest-path trees 
(SPT) and minimum Steiner trees~\cite{moy1994multicast,steiner1991approximation,Hwang1992,Charikar1999}. 
These algorithms aim to minimize the path length or the overall use of the link while
maintaining a tree structure rooted at the source. Despite their simplicity and low 
computational cost, tree-based multicast methods suffer from inherent limitations: 
they rely on a single spanning structure, cannot exploit alternative disjoint paths, 
and exhibit poor adaptability in the presence of link failures or dynamic network
conditions. Extensions such as shared trees and dynamic Steiner 
approximations~\cite{rouskas1997multicast} attempted to address these issues, but 
the fundamental constraints of tree rigidity remained.

The natural formulation of the multicast problem is closely tied to the 
\emph{maximum flow problem}, originally introduced by Ford and Fulkerson~\cite{Ford1956}. 
This framework established the augmenting path method for determining flow capacity
between a source and a sink. Subsequent refinements led to polynomial-time algorithms,
notably Dinic’s blocking flow approach~\cite{Dinic1970} and the Edmonds–Karp algorithm,
which guarantees $O(|V||E|^2)$ complexity by relying on breadth-first searches to
identify augmenting paths~\cite{EdmondsKarp1972}. These algorithms form the theoretical 
foundation for constructing flow-based multicast subgraphs, and they continue to serve 
as baselines for both coding and non-coding multicast strategies.

In contrast, the introduction of network coding~\cite{Ahlswede2000,Li2003,Koetter2003}
revolutionized the multicast problem by demonstrating that coding operations at
intermediate nodes can achieve the maximum information flow simultaneously to all
receivers. Network coding eliminates the bottlenecks imposed by tree-based structures 
and ensures optimal utilization of network capacity. Subsequent work explored 
practical coding strategies, including random linear coding~\cite{chou2003practical},
distributed implementations \cite{ho2006random}, and robust formulations under 
link failures \cite{lun2005network}. Nevertheless, despite their theoretical advantages,
network coding schemes face persistent deployment challenges: (i) coding/decoding 
overhead at intermediate and end nodes, (ii) synchronization and coordination 
complexity across flows, and (iii) incompatibility with existing forwarding hardware 
in traditional IP networks. As a result, network coding has remained primarily within
theoretical and experimental domains, with limited large-scale adoption in operational
networks.

To bridge the gap between tree-based multicast and fully coded approaches, several 
hybrid strategies have been proposed. For example, diversity-aware multicast 
frameworks~\cite{widmer2001extending} leverage multiple disjoint or partially disjoint
trees to improve resilience and throughput. Overlay-based multicast 
systems~\cite{chu2002case,padmanabhan2003resilient} exploit application-level
relays to bypass limitations of native IP multicast, while multipath forwarding 
techniques~\cite{kandula2005walking} balance traffic across parallel paths to 
increase robustness. These methods improve performance compared to single-tree multicast,
but they do not guarantee the optimal throughput achievable by network coding and often
incur additional coordination complexity.

More recently, the concept of \emph{source-coded multicast} has been introduced as a 
viable alternative that retains the efficiency of flow maximization while avoiding
intermediate-node coding. The early work in~\cite{lestayo2001adaptive} already explored 
forward error correction (FEC) for reliable multicast. Building on this foundation, 
subsequent contributions~\cite{Lestayo2023,lestayo2024source} demonstrated how 
source-side processing can enable efficient dissemination by leveraging 
maximum-flow decomposition without relying on traditional tree construction. 
Their approach highlights the potential of source-only coding mechanisms to simplify 
deployment and improve scalability. Our work builds upon this line of research, 
extending it by integrating a systematic overlap management mechanism that enforces 
consistent convergence of receiver paths while maximizing utilization of available 
network capacity.

This paper further extends the source-coded paradigm by introducing 
\emph{overlap-aware online path management}: when multiple overlap events occur, 
a redirection policy at the first intersection enforces loop-free behavior and 
efficient capacity utilization—addressing gaps left by prior approaches.

\section{System Model}
\label{sec:model}

We model the communication network as a directed multigraph
$G=(V,E)$, where $V$ denotes the set of nodes and $E$ the set of directed edges.
Each directed edge $(u,v)\in E$ represents a unit-capacity transmission resource,
\[
c(u,v)=1,
\]
corresponding to one unit of information per time slot. Multiple parallel edges
may exist between the same ordered pair of nodes to model higher-capacity or
redundant connections. The network is assumed to be lossless and error-free at
the packet level; reliability and scheduling aspects are orthogonal to the
throughput analysis considered here.

A single source node $s\in V$ aims to deliver identical content to a set of
receivers $\mathcal{R}=\{r_1,\dots,r_m\}\subset V$. The objective is to maximize
the rate at which all receivers can simultaneously receive the content under
unit-capacity constraints. For each receiver $r_i$, the maximum achievable flow
$f(s,r_i)$ equals the maximum number of edge-disjoint directed paths connecting
$s$ and $r_i$. The overall multicast rate is therefore limited by the receiver
with the smallest individual max-flow:
\begin{equation}
K = \min_i f(s,r_i),
\label{eq:group-maxflow}
\end{equation}
which we refer to as the \emph{multicast group max-flow}.

To compute the path sets $\mathcal{P}_i$ corresponding to each $r_i$, we rely on
classical maximum-flow algorithms. The Ford--Fulkerson
method~\cite{Ford1956} defines the general framework of augmenting paths, while
Dinic’s blocking-flow algorithm~\cite{Dinic1970} and Edmonds--Karp’s refinement
~\cite{EdmondsKarp1972} provide polynomial-time guarantees. Edmonds--Karp is
particularly convenient for the unit-capacity, directed multigraphs considered
here: it identifies augmenting paths using \emph{breadth-first search (BFS)} on
the residual graph, ensuring that the length of augmenting paths is
non-decreasing across iterations and yielding total complexity
$O(|V||E|^2)$.

The residual graph $G_f=(V,E_f)$ is initialized as $E_f\gets E$, so that each
directed edge instance acts as an independent unit-capacity resource. When an
augmenting path uses a directed edge $e=(u,v)$, that arc is saturated and
removed from $E_f$ by setting its residual capacity to zero. The opposite
direction $(v,u)$, if present as a distinct edge, remains available unless it is
also saturated. Bidirectionality is thus modeled explicitly as two separate
arcs, not as a single undirected link.

The Edmonds--Karp procedure for a given receiver $r_i$ is summarized in
Algorithm~\ref{alg:ek-single-integrated}. The algorithm repeatedly performs BFS
on the residual graph to find shortest augmenting paths from $s$ to $r_i$,
backtracks them upon discovery, updates the residual capacities, and stores each
resulting path in $\mathcal{P}_i$. Because all edges have unit capacity, every
augmentation increases the flow by one unit, and all paths in $\mathcal{P}_i$
are mutually edge-disjoint.

\begin{algorithm}[t]
\caption{Edmonds--Karp for $(s,r_i)$}
\label{alg:ek-single-integrated}
\begin{algorithmic}[1]
\Require Graph $G=(V,E)$; source $s$; receiver $r_i$
\Ensure Path set $\mathcal{P}_i$ of edge-disjoint $s{\to}r_i$ paths
\State $\mathcal{P}_i \gets \varnothing$; $E_f \gets E$ \Comment{Initialize residual graph}
\While{true}
    \State $Q \gets [\,s\,]$ \Comment{BFS queue}
    \State $\pi(v) \gets \varnothing,\ \pi_{\text{edge}}(v) \gets \varnothing,\ \forall v \in V$
    \While{$Q \neq [\,]$}
        \State $u \gets \operatorname{head}(Q)$;\quad $Q \gets \operatorname{tail}(Q)$
        \For{each edge $e=(u,v) \in E_f$ with $\pi(v)=\varnothing$}
            \State $\pi(v) \gets u$;\quad $\pi_{\text{edge}}(v) \gets e$;\quad $Q \gets Q \,\Vert\, [\,v\,]$
            \If{$v = r_i$} \Comment{Receiver reached}
                \State $p \gets [\,]$;\quad $x \gets r_i$ \Comment{Backtrack and update $E_f$}
                \While{$x \neq s$}
                    \State $e_x \gets \pi_{\text{edge}}(x)$; $p \gets [\,e_x\,] \,\Vert\, p$
                    \State $c_f(e_x) \gets 0$ \Comment{Saturate edge}
                    \State $x \gets \pi(x)$
                \EndWhile
                \State $\mathcal{P}_i \gets \mathcal{P}_i \cup \{p\}$
                \State \textbf{continue} \Comment{Restart outer loop}
            \EndIf
        \EndFor
    \EndWhile
    \State \textbf{break} \Comment{No augmenting path found}
\EndWhile
\State \Return $\mathcal{P}_i$
\end{algorithmic}
\end{algorithm}

\begin{figure}[t]
  \centering
  \tikzset{every node/.style={font=\footnotesize}}
  \input{figure_ek_steps}
  \caption{Illustrative Edmonds--Karp BFS: (a) layered expansion from $s$;
  (b) predecessor map; (c) reconstruction of the shortest $s{\to}r_i$
  augmenting path.}
  \label{fig:ek-bfs-steps}
\end{figure}
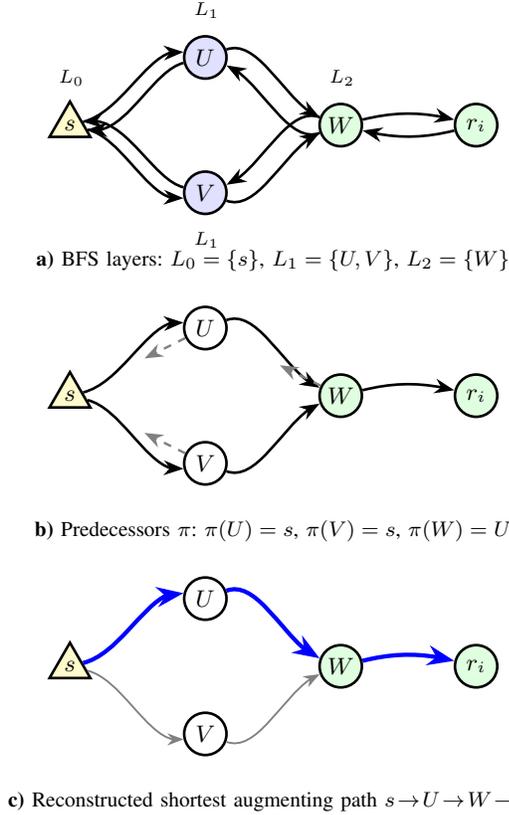

Figure~\ref{fig:ek-bfs-steps} illustrates a single BFS run on an
example topology with intermediate nodes $U$, $V$, and $W$. Panel~(a)
shows the layered expansion of the search from $s$; panel~(b) records the
predecessor map $\pi(\cdot)$ for each discovered node; and panel~(c)
reconstructs the shortest augmenting path by backtracking predecessors,
resulting in $s{\to}U{\to}W{\to}r_i$. Each connected node pair is represented
by two directed arcs, one per direction; the augmenting path uses only one
orientation, leaving the reverse arc available.

Each BFS identifies one additional edge-disjoint path and updates the residual
graph accordingly, as shown in Fig.~\ref{fig:ek-residual}. Because edges have
unit capacity, every augmentation increases the total flow by one, and the set
$\mathcal{P}_i$ provides a complete decomposition of the maximum $s$--$r_i$
flow into unit-capacity paths.

\begin{figure}[t]
  \centering
  \tikzset{every node/.style={font=\footnotesize}} 
  \input{figure_ek}
  \caption{Residual graph before (a) and after (b) one augmentation.
  The arcs used by the shortest $s{\to}r_i$ path are removed,
  while unused and reverse arcs remain available for subsequent searches.}
  \label{fig:ek-residual}
\end{figure}

While this formulation suffices for independent receiver sessions, it does not
directly address the interactions between multiple receivers when their path
sets $\{\mathcal{P}_i\}$ are combined into a single multicast subgraph.
Uncontrolled overlaps across receivers can compromise both capacity efficiency
and the feasibility of source-only coding. To overcome these limitations, the
next section extends Edmonds--Karp into an \emph{online color-constrained BFS},
which incorporates overlap restrictions directly into path discovery.
This online variant ensures that each new receiver flow is integrated
consistently with previously established paths—preserving disjointness whenever
possible and enforcing controlled redirection at overlaps.

\color{black}

\section{Proposed Framework}
\label{sec:framework}

In this Section, we introduce the conceptual foundations of our online multicast 
framework. The goal is to exploit network path diversity while ensuring that 
overlaps between receiver flows remain consistent with source-only coding. 
Unlike classical multicast trees, which rigidly constrain all receivers to a 
single spanning structure, our design integrates multiple edge-disjoint or partially
overlapping paths in a way that is both throughput-efficient and coding-compatible. 
The framework is articulated around three pillars: (i) the motivation and design 
principles that motivate overlap control, (ii) the notion of colors and invariants 
that formalize feasible overlaps, and (iii) the source-coding perspective that ensures 
end-to-end decodability.

\subsection{Motivation and Design Principles}
The motivation for explicitly controlling path overlap arises from prior work on 
source-coded multicast~\cite{Lestayo2023,lestayo2024source}, 
where it was shown that only certain overlap structures between receivers are 
compatible with efficient source-only coding.  In particular, when flows to 
different receivers overlap across multiple distinct paths, the independence 
of the delivered substreams may be compromised, preventing receivers from decoding 
the full content.  On the other hand, if the overlap of a new receiver flow is 
restricted to a single preexisting path, or if it is redirected to consistently
align with that path at the first intersection, the linear independence of the 
substreams is preserved while still compacting the multicast subgraph.  

This principle guides the design of our framework:
\begin{itemize}
\item \emph{Incrementality}: receivers are integrated one by one, without global
recomputation.

\item \emph{Overlap control}: each new path is admitted only if it remains disjoint, 
or if it overlaps consistently with a single existing path.

\item \emph{Feasibility by construction}: redirection at the first overlap ensures 
that infeasible multi-overlaps are automatically avoided.
\end{itemize}
Together, these design principles guarantee that the resulting multicast subgraph is 
both loop-free and compatible with source-only coding.

\subsection{Colors, Zones, and Invariants}
We model the multicast subgraph as $\mathcal{G}_{\text{multicast}}=(V,E_M)$, 
incrementally constructed as receivers are integrated.  Each directed edge $e \in E_M$ 
is labeled with a color from a global palette $\mathcal{C}$.  A color identifies a
\emph{zone}, i.e., a connected subset of edges that originated from one receiver’s 
path and that may later be extended by subsequent receivers aligning with it.  

When a new path is discovered: (i) If it is disjoint from $E_M$, it is assigned a 
fresh color (new zone); (ii) If it overlaps with exactly one existing zone, it 
inherits that zone’s color; (iii) If it encounters multiple overlaps, it is 
automatically redirected at the first intersection to continue within the corresponding
zone, preserving feasibility.

The algorithm preserves the following invariants:
\begin{enumerate}
\item \textbf{Unit-capacity feasibility}: each directed edge is saturated after being 
used by one path, guaranteeing that residual capacity constraints are respected.
\item \textbf{Single-path overlap consistency}: every new path may align with at most 
one preexisting color; multi-overlaps are prevented by redirection.
\item \textbf{Per-receiver exclusivity}: a receiver $r_i$ assigns a distinct color to 
each of its own paths, ensuring independence of its incoming substreams.
\end{enumerate}

\subsection{Source-Coding Perspective}
From a coding perspective, all operations remain confined to the source $s$.  
The source performs flow-splitting into independent substreams, each mapped to a 
unit-capacity path discovered under the color-constrained rules.  Intermediate 
nodes simply forward packets, without any need for recoding.  At the destination, 
each receiver $r_i$ reconstructs the original content by combining the independent
substreams delivered along its disjoint or zone-aligned paths.  

This perspective highlights the advantage of the proposed framework: it achieves much 
of the throughput of network coding, but without requiring complex in-network operations.
Instead, the enforcement of overlap consistency through colors guarantees that the 
source-only coding strategy remains feasible.

In the next Section, we translate these principles into a concrete \emph{online
algorithm}. We formalize the color-constrained breadth-first search, describe the 
path integration process, and illustrate its operation through examples.  This 
provides the step-by-step procedure by which the proposed framework is realized in 
practice.

\section{Online Algorithm}
\label{sec:algorithm}

Building on the system model defined in Section~\ref{sec:model} and the 
design principles outlined in Section~\ref{sec:framework}, we now present 
the online multicast construction algorithm in detail. 
Unlike classical offline approaches, which first compute full maximum-flow 
decompositions for each receiver and then reconcile overlaps, our method 
integrates overlap control directly into the path discovery stage. 
This design allows the multicast subgraph to be constructed incrementally 
as receivers arrive, without global recomputation, while guaranteeing both 
unit-capacity feasibility and consistency of overlaps. 
The algorithm relies on a \emph{color-constrained breadth-first search (BFS)} 
that enforces zone alignment rules during exploration, ensuring that every 
augmenting path is either disjoint, aligned to a single existing zone, 
or redirected at its first intersection. 
We first describe the constrained BFS mechanism, then the integration routine 
for sequential receivers, and finally provide a detailed pseudocode implementation.

\subsection{Color-Constrained BFS}
For each receiver $r_i$, a constrained BFS is executed on the residual 
graph $G_f$. The key difference from standard Edmonds--Karp lies in how 
path expansions are filtered according to the active color state 
of the partial path. Specifically:
\begin{itemize}
  \item If the current partial path has not touched any colored edge, 
  it may traverse any uncolored edge freely. If it enters a colored zone, 
  that color becomes the active color for the remainder of the path.
  \item Once a color $c$ is active, only edges that are either uncolored 
  or already marked with $c$ may be traversed. Expansions that would mix 
  multiple colors are forbidden.
\end{itemize}

This rule has three important consequences:
\begin{enumerate}
  \item Paths that are entirely disjoint from $E_M$ are discovered unchanged 
  and can be added as fresh zones.
  \item Paths that overlap with a single existing zone follow it consistently, 
  thereby inheriting its color and maintaining feasibility.
  \item Paths that would otherwise overlap with multiple zones are 
  automatically redirected at their first intersection, because BFS continues 
  only along the first touched color. This ensures loop-free and 
  overlap-consistent integration without post-processing.
\end{enumerate}
Thus, the BFS not only identifies augmenting paths but also enforces 
the overlap rules by construction.

\subsection{Online Integration Routine}
Receivers are processed sequentially, one at a time. 
For a given receiver $r_i$, the constrained BFS is invoked repeatedly 
until no more admissible augmenting paths can be found. 
Each discovered path is backtracked, assigned a color, and appended to 
the multicast edge set $E_M$. The update rules are as follows:
\begin{enumerate}
  \item \textbf{No overlap (disjoint path)}: a fresh color is drawn from 
  the palette and assigned to all edges in the path. The path forms a 
  new independent zone.
  \item \textbf{Single overlap (aligned path)}: the path inherits the color 
  of the zone it touches, and its edges are appended to $E_M$ under that color.
  \item \textbf{Redirected multi-overlap}: if the BFS encounters multiple 
  zones, it automatically continues along the first touched zone. 
  The resulting path is reconstructed as the prefix up to the intersection 
  plus the existing suffix of the touched zone. This guarantees loop-free 
  redirection and consistent color assignment.
\end{enumerate}

After integration, residual capacities are updated by saturating the 
edges of the discovered path, ensuring that no edge instance is reused 
beyond its unit capacity. 
The routine then attempts another BFS for the same receiver until 
its maximum flow is exhausted, after which the algorithm proceeds 
to the next receiver. 
In this way, the multicast subgraph is constructed online, with 
overlap control embedded in every step.

The complete procedure is summarized in  Algorithm~\ref{alg:online-color-allinone}. 
The pseudocode integrates path discovery, color assignment, residual  updates, 
and receiver-by-receiver processing into a single unified routine. It makes explicit 
the initialization of variables, the BFS state expansion, the backtracking of paths, 
and the incremental updates to both the residual graph and the multicast subgraph. 
This detailed step-by-step description facilitates both theoretical analysis 
and practical implementation of the proposed framework.

\begin{algorithm}[t]
\caption{Online Color-Constrained Multicast Construction}
\label{alg:online-color-allinone}
\begin{algorithmic}[1]
\Require $G=(V,E)$; source $s$; receivers $\{r_1,\dots,r_m\}$
\Ensure Multicast subgraph $G_M=(V,E_M)$ and $\{\mathcal{P}_i\}_{i=1}^m$
\State $E_M \gets \varnothing$;
 $E_{\mathrm{res}} \gets E$ \Comment{Residual Edges}
\State $\mathcal{C} \gets \{c_1,c_2,\dots\}$;\quad $ix \gets 0$ \Comment{Color palette and pointer}
\State $\text{color}:E \to \mathcal{C}\cup\{\varnothing\}$ \Comment{Color function}
\State $\text{color}(e)\gets \varnothing,\ \forall e\in E$ \Comment{All edges start uncolored}
\For{$i=1$ \textbf{to} $m$} \Comment{Process receivers $r_i$}
  \State $\mathcal{P}_i \gets \varnothing$;\quad $C_i \gets \varnothing$;\quad $\mathcal{E}_i \gets \varnothing$ \Comment{Paths, colors, edges}
  \While{\textbf{true}} 
    \State $\textit{prevPaths} \gets |\mathcal{P}_i|$
    \State $Q \gets [\,(s,\varnothing)\,]$ \Comment{BFS queue $(\text{node},\text{active-color})$}
    \State $\pi(v,c)\gets \varnothing,\  \forall v\in V,\ \forall c\in \mathcal{C}\cup\{\varnothing\}$
    \State $\pi_{\text{edge}}(v,c)\gets \varnothing,\ \forall v\in V,\ \forall c\in \mathcal{C}\cup\{\varnothing\}$
    \State $\pi(s,\varnothing)\gets (s,\varnothing)$ \Comment{Source discovered}
    \While{$Q \neq [\,]$}
      \State $(u,c^\star) \gets \operatorname{head}(Q)$;\quad $Q \gets \operatorname{tail}(Q)$
      \For{each $e=(u,v) \in E_{\mathrm{res}}$ \textbf{with} $c_f(e)=1$ \textbf{and} $e \notin \mathcal{E}_i$}
        \State $c_e \gets \text{color}(e)$ \Comment{$c_e\in \mathcal{C}\cup\{\varnothing\}$}
        \If{$c_e \in C_i$} \textbf{continue} \EndIf \Comment{Already used color}
        \If{$c^\star=\varnothing$} \State $c_{\text{next}} \gets c_e$ \Comment{Uncolored}
        \Else
          \If{$c_e \neq \varnothing$ \textbf{and} $c_e \neq c^\star$} \textbf{continue} \EndIf \Comment{Forbid multi-color overlap}
          \State $c_{\text{next}} \gets c^\star$
        \EndIf
        \If{$\pi(v,c_{\text{next}})=\varnothing$} 
          \State $\pi(v,c_{\text{next}}) \gets (u,c^\star)$; $\pi_{\text{edge}}(v,c_{\text{next}}) \gets e$
          \If{$v = r_i$} \Comment{Receiver reached}
          \State \textbf{// Backtrack and update $E_{res}$}
            \State $p \gets [\,]$; $(x,c_x)\gets (r_i,c_{\text{next}})$ 
            \While{$x \neq s$}
              \State $e_x \gets \pi_{\text{edge}}(x,c_x)$; $p \gets [\,e_x\,] \,\Vert\, p$
              \State $(x,c_x) \gets \pi(x,c_x)$
            \EndWhile
            \State $\Omega(p) \gets \{\, \text{color}(e) : e \in p,\ \text{color}(e)\neq\varnothing \,\}$
            \If{$|\Omega(p)|=0$} \Comment{No overlap}
              \State $ix \gets ix+1$; $c \gets c_{ix}$ 
            \Else 
            \State $c \gets \operatorname{Only}(\Omega(p))$
            \EndIf
             \State  $\text{color}(e)\gets c, \forall e\in p$
             \State $E_M \gets E_M \cup p$; $\mathcal{P}_i \gets \mathcal{P}_i \cup \{p\}$
             \State $C_i \gets C_i \cup \{c\}$; $\mathcal{E}_i \gets \mathcal{E}_i \cup p$
            \For{each $e\in p$}  $c_f(e)\gets 0$ \EndFor \Comment{Residual update}
            \State \textbf{break} \Comment{Path found}
          \Else \quad $Q \gets Q \,\Vert\, [\,(v,c_{\text{next}})\,]$
          \EndIf
        \EndIf
      \EndFor
      \If{$|\mathcal{P}_i|>\textit{prevPaths}$}  \textbf{break} \EndIf 
    \EndWhile
    \If{$|\mathcal{P}_i|=\textit{prevPaths}$} 
      \textbf{break} \Comment{BFS exhausted}
    \EndIf
  \EndWhile
\EndFor
\State \Return $G_M=(V,E_M)$ and $\{\mathcal{P}_i\}_{i=1}^m$
\end{algorithmic}
\end{algorithm}

\subsection{Algorithm Walkthrough}

For clarity, we now describe the main components of
Algorithm~\ref{alg:online-color-allinone} step by step, grouped by line ranges.

\textbf{Lines 1--4 (Initialization).}
The algorithm begins by initializing the multicast edge set $E_M$ and the residual
edge set $E_{\mathrm{res}}$, which contains all directed unit-capacity edges that
remain unsaturated. A global color palette $\mathcal{C}=\{c_1,c_2,\dots\}$ is
defined together with a pointer $ix$ to allocate fresh colors. The function
$\text{color}(\cdot)$ maps each edge to its assigned color, and initially all edges
are uncolored.

\textbf{Lines 5--6 (Receiver loop).}
Receivers are processed sequentially. For each receiver $r_i$, we maintain three sets:
$\mathcal{P}_i$ (the set of paths integrated for $r_i$),
$C_i$ (the colors already used by $r_i$ to ensure one color per path),
and $\mathcal{E}_i$ (the edges already reserved for $r_i$ to guarantee
edge-disjointness of its paths).
The outer while-loop attempts to find multiple admissible $s{\to}r_i$ paths
until none remain.

\textbf{Lines 7--13 (BFS setup).}
A new breadth-first search (BFS) is initialized from $(s,\varnothing)$,
where the state includes both the current node and the active color.
The predecessor maps $\pi(\cdot,\cdot)$ and $\pi_{\text{edge}}(\cdot,\cdot)$
are cleared, and the source is marked as discovered.
The variable $\textit{prevPaths}$ records the number of paths found so far for $r_i$
and will be used to determine whether the BFS successfully augmented the flow.

\textbf{Lines 14--29 (BFS exploration).}
Nodes are extracted from the queue one by one.
For each outgoing residual edge $e=(u,v)$ with capacity one and not already
used by $r_i$, the algorithm evaluates its color $c_e$.
Two constraints are enforced:
(i) a receiver cannot reuse a color it has already consumed across its paths
($c_e \notin C_i$), and
(ii) if the path is already following a color $c^\star$,
then only edges that are uncolored or matching $c^\star$ may be traversed
(multi-color overlap is forbidden).
The resulting color state $c_{\text{next}}$ is attached to the successor
state $(v,c_{\text{next}})$.
If this state has not been visited, it is discovered and linked through the
predecessor maps.

\textbf{Lines 30--49 (Path reconstruction and integration).}
If the sink $r_i$ is reached, the shortest admissible augmenting path
is reconstructed by backtracking predecessors from $(r_i,c_{\text{next}})$
to $(s,\varnothing)$.
The set $\Omega(p)$ collects the colors already present on the path:
\begin{itemize}
\item If $\Omega(p)=\varnothing$, the path introduces no overlap,
and a fresh color $c_{ix+1}$ is assigned.
\item Otherwise, by construction, $\Omega(p)$ contains exactly one color,
which is reused consistently for the entire path.
\end{itemize}
All edges of $p$ are assigned the selected color $c$,
added to $E_M$, and appended to $\mathcal{P}_i$.
The color $c$ and the edges of $p$ are also recorded in $C_i$ and $\mathcal{E}_i$.
Finally, the residual graph is updated by saturating each used edge
($c_f(e)\gets 0$), ensuring that the same edge cannot be reused by another path.
After this, the BFS terminates and the outer while-loop resumes to attempt
another augmentation for $r_i$.

\textbf{Lines 50--56 (Termination for receiver $r_i$).}
If no new path was discovered in the last BFS run
($|\mathcal{P}_i|=\textit{prevPaths}$),
the while-loop ends, meaning the maximum flow to $r_i$
under the color constraints has been reached.
The algorithm then proceeds to the next receiver.

\textbf{Final return.}
After all receivers have been processed, the algorithm outputs the multicast
subgraph $G_M=(V,E_M)$ together with the per-receiver path sets
$\{\mathcal{P}_i\}_{i=1}^m$.

\begin{figure}[t]
  \centering
 \input{online_example}
  \caption{Step-by-step online integration of three receivers: 
  (a) disjoint paths for $r_1$ (blue, green); 
  (b) integration of $r_2$ with one disjoint path (red) and one single overlap (green); 
  (c) integration of $r_3$ with redirection at the first overlap (blue).}
  \label{fig:online-example}
\end{figure}

\color{black}
\subsection{Illustrative Example}
\label{sec:examples}

To illustrate the operation of the proposed online algorithm, we now 
present a step-by-step example on an illustrative topology. 
The goal is not to represent a large-scale network, but rather to make 
visible how the color-constrained BFS and the online integration rules 
interact when multiple receivers are sequentially added. 
For clarity, all links in the figures are drawn in gray, and the active 
augmenting paths discovered by the algorithm are highlighted in color. 
Different colors represent distinct flow zones created during the process. 

\subsubsection{First Receiver}
We begin with receiver $r_1$, assumed to have a min-cut of two with respect 
to the source $s$. The first BFS run discovers a path 
$s \to U \to W \to r_1$, which is disjoint and therefore assigned a fresh color 
(blue). A second BFS identifies an edge-disjoint path 
$s \to V \to r_1$, assigned another fresh color (green). 
At this point, the multicast subgraph $E_M$ for $r_1$ consists of two 
independent zones, guaranteeing that $r_1$ receives two linearly 
independent substreams. Figure~\ref{fig:online-example}~(a) shows this stage.

\subsubsection{Second Receiver}
Next, receiver $r_2$ is integrated. The first BFS discovers a path 
$s \to U \to r_2$, which is completely disjoint from the existing zones; 
it is therefore assigned a fresh color (red). A second BFS then identifies 
a path $s \to V \to W \to r_2$, which touches the preexisting green zone at 
edge $(s,V)$ and consistently follows it. This path is thus integrated into 
the green zone. As a result, $r_2$ receives two independent substreams, 
one red and one green. Figure~\ref{fig:online-example}~(b) illustrates this 
integration.

\subsubsection{Third Receiver}
Finally, receiver $r_3$, also with min-cut two, is processed. 
The first BFS identifies a disjoint path $s \to X \to r_3$, which does not 
intersect any existing zone. This creates a fresh zone, highlighted in 
orange. A second BFS then finds a candidate path through $W$ that would 
intersect multiple zones. Instead of allowing a multi-overlap, the algorithm 
redirects the suffix at the first touched zone (blue), forcing the path to 
align entirely with it. The resulting path therefore consists of a blue-colored 
suffix from $W \to r_3$, ensuring feasibility and preventing cycles. At this 
stage, $r_3$ receives two substreams: one disjoint (orange) and one aligned 
with an existing zone (blue). Figure~\ref{fig:online-example}~(c) illustrates 
this final integration.

\subsection{Discussion}
These examples highlight how the online algorithm adapts naturally to 
successive receiver integrations. Each new receiver is guaranteed to 
receive the maximum number of admissible substreams permitted by the 
network min-cut, while overlaps are managed in a way that is consistent 
with source-only coding. The use of colors makes the invariants 
transparent: each receiver obtains substreams of distinct colors, 
no path uses edges of more than one color, and redirections occur 
automatically at the first overlap. 

This illustrative topology also emphasizes the scalability advantage of the 
online approach: each integration is performed incrementally without 
recomputing global flows. In the next section, we formalize this 
intuition by analyzing the algorithm’s complexity and comparing it 
with classical Edmonds--Karp and coding-optimal benchmarks.

\section{Theoretical Analysis}
\label{sec:theoretical}

\subsection{Correctness}
The proposed algorithm admits a theoretical characterization in terms of 
correctness, feasibility, and its relation to classical notions of maximum flow 
and multicast construction. The following discussion highlights its fundamental
properties.

The algorithm guarantees loop-free integration of receiver flows. By construction, 
every newly added path either introduces a disjoint flow or aligns with a single 
existing flow upon its first overlap. As a result, each directed edge instance 
is saturated at most once, ensuring strict adherence to unit-capacity feasibility.
Furthermore, the breadth-first nature of the search ensures that augmenting paths
are the shortest admissible ones in hop count, consistent with the classical 
Edmonds--Karp framework.

From the perspective of throughput, the online construction is always bounded above 
by the maximum-flow value $f(s,r_i)$ for each receiver. In the absence of multi-path
overlaps, this bound is achieved exactly. When redirection occurs, the resulting path 
set may fall short of $|f(s,r_i)|$ disjoint paths, but the loss is limited in practice.
In our ER/WS experiments we did not observe a degradation of the group multicast 
max-flow when comparing unconstrained EK with its restricted variant (cf. 
Section~\ref{sec:results}). Nonetheless, this near-optimality is not universal: 
there exist small, carefully crafted unit-capacity digraphs where the overlap 
constraint strictly reduces the achievable multicast throughput. 
Figure~\ref{fig:adversarial_restricted_loss} provides a minimal counterexample: 
per-receiver EK finds two edge-disjoint paths to every receiver, whereas the 
restricted integration forces a first-touch alignment that prevents two independent 
zones from serving all receivers simultaneously, thereby reducing the multicast rate.

A key structural property of the algorithm is its treatment of overlaps. By enforcing 
that overlaps are confined to a single preexisting path, the resulting multicast 
subgraph is naturally partitioned into edge-disjoint zones. This structure has two
important implications. First, it ensures compatibility with source coding:
each zone can be associated with a distinct substream, enabling receivers to 
reconstruct the transmitted content without requiring intermediate coding operations.
Second, it yields a compact, tree-like topology: although the subgraph is not a tree 
in the strict sense, redirected flows converge consistently toward the source,
producing a structure with branching behavior analogous to multicast trees.

\begin{figure}[t]
  \centering
  \input{restrictedvsunrestricted}
  \caption{Adversarial topology showing a strict loss under overlap-constrained integration.
  (a) Topology only. (b) Per-receiver EK: two edge-disjoint paths to every receiver.
  (c) Restricted EK: the first-touch overlap rule collapses one branch for $r_3$,
  reducing the multicast group throughput.}
  \label{fig:adversarial_restricted_loss}
\end{figure}

When compared to alternative multicast approaches, the proposed framework occupies 
an intermediate position. Relative to tree-based multicast, it achieves higher 
resilience by preserving multiple flow diversity paths where available. Relative to 
fully coded multicast, it sacrifices only marginal throughput while avoiding the
operational complexity of in-network coding. Thus, the algorithm combines the
feasibility and efficiency of Edmonds--Karp with structural guarantees that make it 
source-coding compatible and practically deployable at scale. The previous analysis
demonstrated that the proposed online construction yields a loop-free, capacity-feasible
multicast subgraph that approaches the maximum attainable group flow. The next step 
is to specify how the source uses this structure to assign information to its outgoing
links, thereby realizing source-only coding across the colored multicast tree.

\subsection{Complexity Analysis}
\label{sec:complexity}

The online multicast algorithm inherits its fundamental complexity from Edmonds--Karp,
while introducing lightweight color checks that do not alter the asymptotic order. 
Let $|E|$ denote the number of directed edge instances in the multigraph (counting
parallels). For a fixed receiver $r_i$, each constrained breadth-first search (BFS) 
runs in $O(|E|)$ time. Since all edges have unit capacity, every augmentation 
increases the flow by exactly one, and at most $O(|E|)$ augmentations can be performed.
Hence, the complexity per receiver session is $O(|E|^2)$. Processing all $m$ 
receivers sequentially yields an overall complexity of $O(m|E|^2)$. The additional 
color checks performed during BFS expansion are constant-time operations, so they do not 
modify this bound.

For fairness, these figures correspond exclusively to the construction of multicast
topologies. Classical Edmonds--Karp applied independently to each receiver without 
overlap restrictions has the same $O(m|E|^2)$ cost, but requires additional 
reconciliation of overlapping paths after the fact. Tree-based heuristics, such as
shortest-path or Steiner approximations, typically range between $O(|E|\log|V|)$ 
and $O(|V||E|)$ in complexity, and are therefore computationally lighter, but 
structurally limited since they cannot exploit multiple edge-disjoint paths and thus 
fall short of the throughput achieved by flow-based methods.

In comparison, constructing network coding–optimal multicast subgraphs also requires
computing maximum flows for each receiver, leading again to $O(m|E|^2)$ complexity at 
the topology stage. However, full network coding adds an additional layer of processing:
encoding at the source and decoding at the receivers involve solving systems of linear
equations whose dimension scales with the number of receivers and active subflows,
resulting in at least $O(m^3)$ computational cost, not accounting for per-packet 
operations at intermediate nodes. By contrast, the source-coding paradigm enabled by 
our online algorithm requires only lightweight linear combinations at the source
and no in-network coding, which drastically reduces implementation complexity
while preserving near-optimal throughput.

\section{Source-Only Information Encoding}
\label{sec:sourceassignment}

The online algorithm described in Sections~\ref{sec:framework}--\ref{sec:theoretical}
constructs a multicast subgraph $G_M=(V,E_M)$ that satisfies the structural constraints
required for source-only coding: each receiver $r_i$ is connected to the source $s$
through $k_i$ edge-disjoint or zone-aligned paths, and every path of a receiver is
assigned a distinct color. This Section completes the framework by defining the 
source-side rule that determines what information is transmitted on each outgoing edge 
of $s$, thus achieving the multicast group maximum flow without in-network recoding.

Once the online algorithm has produced the colored multicast subgraph, each color
identifies a distinct transmission branch originating at the source. Different colors 
may reach disjoint or overlapping subsets of receivers, and the total number of colors
$|\mathcal{Z}|$ is generally greater than or equal to the multicast max-flow value.
Hence, the colored structure defines the potential information outlets from the source, 
but not yet the specific symbols to be transmitted on them. At this stage, the 
symbol-assignment procedure proposed by~\cite{Lestayo2023,lestayo2024source} is 
applied to determine, for each color, whether the transmitted content should be a
clear (uncoded) symbol or a linear combination of source symbols, depending on the 
subset of receivers reached by that color. For instance, if the maximum multicast flow 
is $K=1$ and two colors depart from the source to reach disjoint receiver subsets,
both colors carry the same clear symbol, with no coding required. In contrast, when
receivers share parts of the multicast tree, the source assigns coded symbols across 
the corresponding colors to preserve the linear independence of the information 
observed at each receiver.

Following the formulation adopted throughout this paper, the multicast group throughput 
is limited by the receiver with the smallest individual max-flow:
\begin{equation}
K = \min_i f(s,r_i) = \min_i k_i,
\label{eq:groupflow}
\end{equation}
so that all receivers can decode the same number of independent symbols.
At each time slot, the source generates $K$ base symbols $X = [x_1, x_2, \ldots, x_K]^T$,
where each $x_j$ belongs to a finite field $\mathbb{F}_q$. Each outgoing edge 
$e=(s,v)$ is associated with a color $z(e)\in\mathcal{Z}$, and the symbol 
transmitted along it is a linear combination
\begin{equation}
y_e = \mathbf{g}_e^T X,
\label{eq:sourcecombination}
\end{equation}
where $\mathbf{g}_e\in\mathbb{F}_q^K$ is a coding vector drawn from a global source 
matrix $G$. The set of vectors $\{\mathbf{g}_e : e\in E_M,\ s\in e\}$ is chosen to 
be linearly independent, and all edges sharing the same color transmit identical 
mixtures, i.e.\ $y_e=y_{e'}$ whenever $z(e)=z(e')$. Hence, the color-to-vector mapping 
$\{z\mapsto\mathbf{g}_z\}$ is defined once at the source and reused consistently along 
the entire zone.

Each receiver $r_i$ observes the symbols corresponding to the set of colors
$Z_i\subseteq\mathcal{Z}$ that reach it, forming $Y_i = G_{Z_i,:}\, X$.
Receiver $r_i$ successfully decodes the $K$ source symbols if the submatrix 
$G_{Z_i,:}$ has full rank $K$, i.e.
\begin{equation}
\operatorname{rank}\!\big(G_{Z_i,:}\big) = K.
\label{eq:rankcondition}
\end{equation}
This condition is guaranteed if $G$ is constructed as a Vandermonde or Cauchy matrix 
over a field of size $q\ge|\mathcal{Z}|$, since any subset of at most $K$ rows is 
linearly independent. Consequently, all receivers obtain $K=\min_i k_i$ 
independent combinations, matching the multicast group max-flow defined in~\eqref{eq:groupflow}.

To illustrate, consider a source connected to three receivers through three outgoing
zones. With $K=2$ and base symbols $A$ and $B$, the source transmits $y_{z_1}=A$,
$y_{z_2}=B$, and $y_{z_3}=A{+}B$. Receiver $r_1$ receives $\{A,B\}$, $r_2$ receives 
$\{A,A{+}B\}$, and $r_3$ receives $\{B,A{+}B\}$. Each receiver obtains two linearly
independent combinations of $\{A,B\}$, thus recovering all symbols and achieving the 
group throughput $K=2$. This example reproduces the principle proposed 
by~\cite{Lestayo2023,lestayo2024source}, generalized here to the online multicast 
topology produced by our framework.

A practical realization of the matrix $G$ can be obtained using Reed--Solomon (RS)
encoding over $\mathbb{F}_q$. In this case, the $K$ source symbols correspond to 
the information symbols of an $(N,K)$ MDS block code, and the set of $N=|\mathcal{Z}|$ c
coded symbols $\{y_z\}$ is produced by evaluating the associated RS generator 
polynomial at $N$ distinct field elements $\{\alpha_z\}$. This construction ensures 
that any subset of $K$ coded symbols is linearly independent, thereby satisfying 
condition~\eqref{eq:rankcondition} by design. In practice, the source assigns one 
coded symbol $y_z$ to each color and injects it into the network according to the
mapping established by the online algorithm. Since intermediate nodes merely forward 
packets, the resulting scheme realizes source-only linear coding with the same 
algebraic guarantees as the Reed--Solomon code, while remaining fully compatible with 
the topology produced by the color-constrained construction.

In summary, once the multicast subgraph satisfies the single-overlap invariant, the 
source can directly map each color to a predetermined coding vector and transmit 
the corresponding linear combinations according to~\eqref{eq:sourcecombination}.
Intermediate nodes simply forward packets of their assigned color, with no recoding 
or coordination required. Together with the online construction algorithm, this 
source-side information assignment provides a complete, constructive realization
of the source-coded multicast paradigm, achieving the maximum feasible multicast
throughput with minimal operational complexity.

\section{Performance Evaluation}
\label{sec:results}

\subsection{Simulation Setup}
To evaluate the performance of the proposed framework, we carried out
simulations on two widely used random graph models: the Erdős--Rényi 
(ER) model~\cite{Erdos1959} and the Watts--Strogatz (WS) small-world
model~\cite{Watts1998}. These models have been extensively employed in 
the networking literature to capture different structural properties: ER 
graphs provide a baseline of uniform randomness, while WS graphs incorporate 
both clustering and short average path length, thus better reflecting 
realistic communication networks.

For both ER and WS models, the total number of nodes was varied from
$10$ to $200$ in increments of $5$. The receiver set was chosen randomly
with densities between $5\%$ and $25\%$, in steps of $5\%$. Likewise,
the link density was explored in the range $10\%$ to $50\%$, also in
steps of $5\%$. All edges were assigned unit capacity, and for each
parameter setting multiple random graph realizations were generated;
results were averaged to ensure statistical significance. For the WS
model, the rewiring probability was fixed at $\beta=0.1$, while in the ER
model no such parameter is defined. In both cases, the expected mean
degree is consistent with the chosen link density.

\begin{figure}[t]
  \centering
  \begin{subfigure}{\columnwidth}
    \centering
    \includegraphics[width=0.8\linewidth]{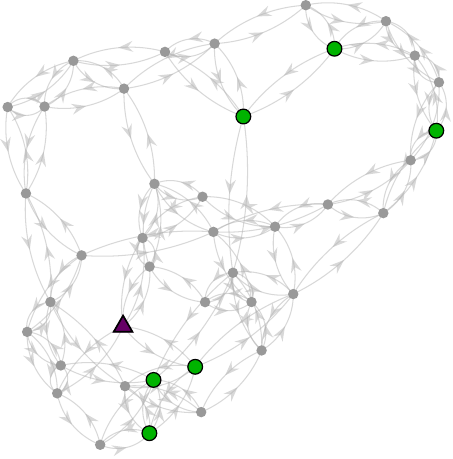}
    \caption{Original WS graph: source (triangle) and receivers (green circles).}
    \label{fig:grafo_original_ws}
  \end{subfigure}


  \begin{subfigure}{\columnwidth}
    \centering
    \includegraphics[width=0.8\linewidth]{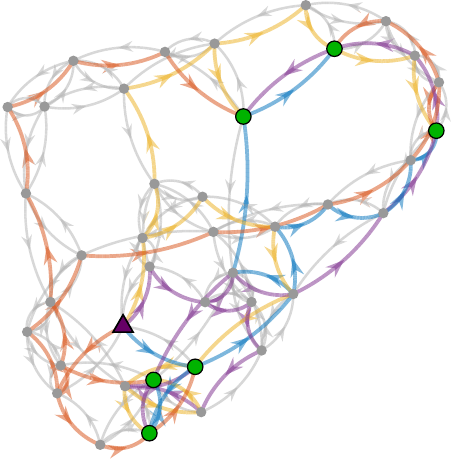}
    \caption{Online multicast construction for WS.}
    \label{fig:grafo_online_ws}
  \end{subfigure}

  \caption{Illustrative Watts--Strogatz example ($n{=}40$, $k{=}4$, $\beta{=}0.1$).}
  \label{fig:grafos_ws}
\end{figure}
\begin{figure}[t]
  \centering
  \begin{subfigure}{\columnwidth}
    \centering
    \includegraphics[width=0.8\linewidth]{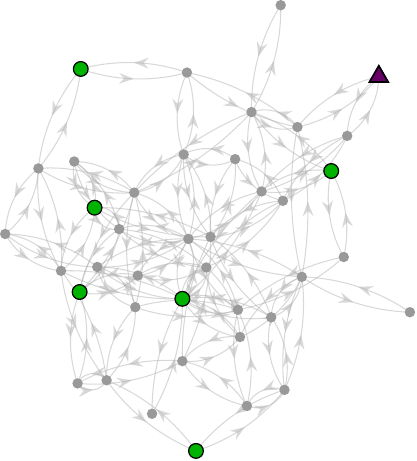}
    \caption{Original ER graph: source (triangle) and receivers (green circles).}
    \label{fig:grafo_original_er}
  \end{subfigure}


  \begin{subfigure}{\columnwidth}
    \centering
    \includegraphics[width=0.8\linewidth]{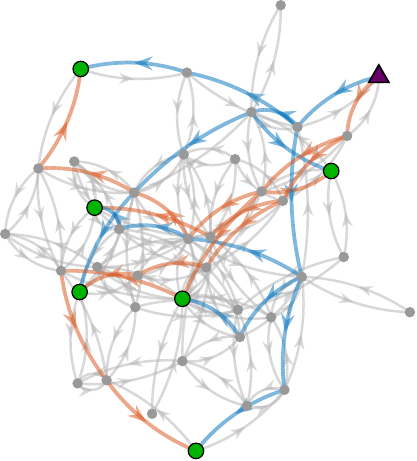}
    \caption{Online multicast construction for ER.}
    \label{fig:grafo_online_er}
  \end{subfigure}

  \caption{Illustrative Erdős--Rényi example ($n{=}40$, $p{\approx}0.103$).}
  \label{fig:grafos_er}
\end{figure}

To build intuition, Figs.~\ref{fig:grafos_ws} and~\ref{fig:grafos_er}
show representative instances with $40$ nodes, $6$ receivers, and mean
degree $k=4$. The WS example (Fig.~\ref{fig:grafos_ws}) corresponds to
$\beta=0.1$, while the ER example (Fig.~\ref{fig:grafos_er}) achieves the
same expected degree with $p \approx 0.103$ (equivalently $m=80$ edges).
In both cases, panel (a) shows the original topology---with the source
marked as a triangle and the receivers highlighted in green---and panel
(b) depicts the corresponding online construction of the multicast
subgraph. Together, these examples illustrate how both graph models can
be instantiated under equivalent parameters, enabling a fair comparison
of the proposed framework.

Finally, it is worth noting that, although the multicast max-flow towards
the group remains unchanged, the restricted Edmonds--Karp variant may
reduce the maximum flow observed at individual receivers.

\subsection{Throughput Results}

\begin{figure}[t]
  \centering
  \input{performance_1_WS}
  \label{fig:adversarial_restricted_loss_WS}
\end{figure}

\begin{figure}[t]
  \centering
  \input{performance_1_ER}
  \label{fig:adversarial_restricted_loss_ER}
\end{figure}

\begin{figure}[t]
  \centering
  \input{performance_2}
  \label{fig:ws_k4_highrec_scaling}
\end{figure}

Across our main grid of ER and WS configurations, the online construction
delivers identical multicast throughput to the offline benchmark
obtained by running Edmonds--Karp independently for each receiver. This
holds throughout the ranges considered for network size, edge density,
and receiver density. Figures~\ref{fig:adversarial_restricted_loss_WS}
and~\ref{fig:adversarial_restricted_loss_ER} report the aggregate multicast
throughput as the receiver density $\rho$ increases, confirming the
coincidence between the restricted online construction and the
unconstrained EK baseline across both random graph models.

A closer look at these figures reveals consistent scaling patterns across
both topologies. For a fixed link density, the total multicast throughput
increases with the network size $n$. This trend is expected since, under
constant edge density, the absolute number of links grows quadratically
with $n$, providing more potential disjoint routes between the source and
the receivers. Similarly, higher link densities yield proportionally
larger throughput values across all receiver densities, reflecting the
linear dependence between aggregate capacity and connectivity.

When comparing the two models, WS graphs exhibit slightly steeper
throughput curves than ER for equivalent link and receiver densities.
This difference arises from the small-world structure of the WS model:
its higher clustering and shorter characteristic path length enhance the
availability of alternative low-hop routes, thus increasing the effective
diversity of disjoint paths. In contrast, the ER model ---lacking such
structural shortcuts--- shows smoother growth with $n$ and lower marginal
gains as receiver density increases. Nonetheless, both models confirm the
same qualitative trends and, most importantly, the same quantitative
match between the restricted and unconstrained Edmonds--Karp outcomes
across the examined parameter space.

To further investigate the limits of this correspondence, we conducted
a set of stress tests designed to exacerbate potential overlap conflicts
in the online construction. Specifically, we fixed the mean degree to
$k=4$ in the WS model (with $\beta=0.1$) and progressively scaled the
network size $n$ while imposing a dense receiver configuration
($r/n=30\%$). Holding $k$ constant isolates the effect of size under
sparse per-node capacity: as $n$ increases, the number of directed
edges grows linearly ($m=nk$), while the edge density
$\rho=m/m_{\max}\approx k/(n{-}1)$ decays inversely with $n$. In
parallel, the receiver set expands proportionally, amplifying fan-out
pressure near the source and increasing the likelihood of path overlap.
This regime is therefore the most adversarial for the restricted
integration rule. As shown in Fig.~\ref{fig:ws_k4_highrec_scaling}, 
the restricted construction continues to match the unconstrained EK across 
almost all sizes; the \emph{first} deviation appears at $n=840$, with a
maximum gap of only one flow unit thereafter.

These stress-test observations reinforce the theoretical expectations:
although adversarial configurations can enforce a strict loss under the
overlap constraint, such structures are extremely rare within ER or WS
ensembles. Empirically, even under dense receiver sets and sparse degree,
throughput losses remain infrequent, small in magnitude, and bounded by 
a single flow unit. This outcome demonstrates that the proposed online
integration rule preserves near-optimal throughput in practically all 
random topologies, validating its efficiency and robustness as a 
scalable multicast construction method. After confirming that the proposed 
online framework matches the optimal throughput of the Edmonds--Karp 
benchmark, we now turn to its computational aspects. The following 
analysis examines how runtime scales with the number of nodes, link density, 
and receiver population in both ER and WS networks.

\subsection{Scalability}
We evaluate scalability by measuring runtime as a function of network
size, receiver density, and average degree for both ER and WS ensembles.
Results consistently show that execution time grows approximately
quadratically with the number of edges, in line with the expected
$O(|E|^2)$ bound per receiver. However, the proposed online construction
achieves substantial practical gains by eliminating redundant
recomputations and incrementally reusing previously integrated flow
structure.

Across all tested configurations, the online algorithm maintains the
same multicast throughput as the reference Edmonds--Karp baseline while
progressively reducing runtime as network connectivity increases.
Higher link densities, which correspond to larger average node degree,
yield proportionally shorter execution times relative to EK. This effect
stems from the online method’s ability to reuse colored paths and merge
zones efficiently: denser topologies provide more alternative disjoint
routes, lowering the number of residual searches and augmentations
required per receiver.

Figure~\ref{fig:runtime} illustrates this behavior for the
Watts--Strogatz (WS) model with rewiring probability $\beta=0.1$.
The plot reports the runtime ratio between the proposed online algorithm
and the per-receiver Edmonds--Karp baseline as the number of nodes $n$
increases. The dashed line marks parity ($=1$), and values below it
indicate faster execution of the online algorithm. The ratio decreases
monotonically with link density, confirming that efficiency improves as
connectivity grows. Equivalent trends were observed for Erdős--Rényi (ER) 
graphs under the same parameter settings, indicating that the observed 
scalability improvement is consistent across both random graph models.

The improvement is particularly pronounced in small-world structures,
where high clustering and short characteristic path lengths favor
path reuse and further reduce exploration overhead. Overall, empirical
results confirm that the online construction scales gracefully with
network size and density: it preserves optimal multicast throughput
while achieving significant and increasing computational efficiency as
the underlying connectivity grows.

\begin{figure}[t]
  \centering
  \input{runtime_ratio.tex}
  \label{fig:runtime}
\end{figure}

\section{Conclusion and Future Work}
\label{sec:conclusion}

This paper introduced an online source-coded multicast framework that 
incrementally integrates receiver flows while explicitly controlling 
path overlaps. By embedding color-constrained breadth-first search into 
the path discovery process, the algorithm guarantees that each new flow 
is either disjoint, aligned with a single preexisting flow, or redirected 
at its first intersection. This construction preserves unit-capacity 
feasibility, enforces overlap consistency, and ensures compatibility 
with source-only coding ---all without requiring global recomputation. 

Our theoretical analysis established that the algorithm operates in 
$O(|E|^2)$ time per receiver, on par with classical Edmonds--Karp, 
while enforcing structural invariants absent in offline approaches. 
Extensive simulations on Erdős–Rényi and Watts–Strogatz random networks 
demonstrated that, in practice, the online construction consistently 
matches the maximum-flow throughput benchmark, confirming its robustness 
and efficiency. Moreover, the resulting multicast subgraphs offer 
enhanced resilience and path diversity compared to tree-based methods, 
while avoiding the complexity of intermediate-node coding inherent to 
network coding.

\textbf{Future work.} Several directions remain open.  First, while no 
throughput degradation was observed in random networks,  structured 
topologies such as scale-free (Barabási–Albert) or stochastic block 
models may reveal conditions where the online  constraints limit 
achievable flow.  Second, further analysis of resilience under dynamic 
conditions  (e.g., receiver churn, link failures) would strengthen the 
practical  relevance of the approach.  Third, integrating lightweight 
coding extensions at the source (e.g., adaptive redundancy or hybrid 
substream designs) may improve robustness to packet loss while preserving 
the no-recoding property at intermediate  nodes.  Finally, applying the 
framework to real-world network traces and  deployments would provide 
additional validation of its scalability  and implementability. Overall, 
the proposed online algorithm provides a practical middle ground  between 
tree-based multicast and full network coding, offering a  scalable, 
source-only solution for next-generation group communication  services.

\bibliographystyle{IEEEtran}
\bibliography{SourceCoding}
\end{document}

%% file: figure_ek_steps.tex
\begin{tikzpicture}[scale=.9,>=Stealth,
  node/.style={circle,draw=black,line width=1pt,minimum size=16pt,inner sep=1pt}]

\node[regular polygon,regular polygon sides=3,draw=black,line width=1pt,
      minimum size=16pt,inner sep=1pt,fill=yellow!25] (s1) at (0,3) {\small $s$};
\node[node,fill=blue!12]   (U1) at (2,4) {\small $U$};
\node[node,fill=blue!12]   (V1) at (2,2) {\small $V$};
\node[node,fill=green!12]  (W1) at (4,3) {\small $W$};
\node[node,fill=green!12]  (r1) at (6,3) {\small $r_i$};

\draw[->,line width=1pt] (s1) .. controls +(0.8,0.2) and +(-0.8,0.2) .. (U1);
\draw[->,line width=1pt] (U1) .. controls +(-0.8,-0.2) and +(0.8,-0.2) .. (s1);

\draw[->,line width=1pt] (s1) .. controls +(0.8,-0.2) and +(-0.8,-0.2) .. (V1);
\draw[->,line width=1pt] (V1) .. controls +(-0.8,0.2) and +(0.8,0.2) .. (s1);

\draw[->,line width=1pt] (U1) .. controls +(0.8,0.3) and +(-0.8,0.3) .. (W1);
\draw[->,line width=1pt] (W1) .. controls +(-0.8,-0.3) and +(0.8,-0.3) .. (U1);

\draw[->,line width=1pt] (V1) .. controls +(0.8,-0.3) and +(-0.8,-0.3) .. (W1);
\draw[->,line width=1pt] (W1) .. controls +(-0.8,0.3) and +(0.8,0.3) .. (V1);

\draw[->,line width=1pt] (W1) .. controls +(0.8,0.2) and +(-0.8,0.2) .. (r1);
\draw[->,line width=1pt] (r1) .. controls +(-0.8,-0.2) and +(0.8,-0.2) .. (W1);

\node[anchor=west] at (-0.3,3.7) {\scriptsize $L_0$};
\node[anchor=west] at (1.7,4.7) {\scriptsize $L_1$};
\node[anchor=west] at (1.7,1.3) {\scriptsize $L_1$};
\node[anchor=west] at (3.7,3.7) {\scriptsize $L_2$};

\node at (3,1) {\footnotesize \textbf{a)} BFS layers: $L_0=\{s\}$, $L_1=\{U,V\}$, $L_2=\{W\}$} ;

\begin{scope}[yshift=-4cm]
\node[regular polygon,regular polygon sides=3,draw=black,line width=1pt,
      minimum size=16pt,inner sep=1pt,fill=yellow!25] (s2) at (0,3) {\small $s$};
\node[node] (U2) at (2,4) {\small $U$};
\node[node] (V2) at (2,2) {\small $V$};
\node[node,fill=green!12] (W2) at (4,3) {\small $W$};
\node[node,fill=green!12] (r2) at (6,3) {\small $r_i$};

\draw[->,line width=1pt] (s2) .. controls +(0.8,0.2) and +(-0.8,0.2) .. (U2);
\draw[->,line width=1pt] (s2) .. controls +(0.8,-0.2) and +(-0.8,-0.2) .. (V2);
\draw[->,line width=1pt] (U2) .. controls +(0.8,0.3) and +(-0.8,0.3) .. (W2);
\draw[->,line width=1pt] (V2) .. controls +(0.8,-0.3) and +(-0.8,-0.3) .. (W2);
\draw[->,line width=1pt] (W2) .. controls +(0.8,0.2) and +(-0.8,0.2) .. (r2);

\draw[gray,dashed,->,line width=.9pt] (U2) -- ($(s2)!0.55!(U2)$);
\draw[gray,dashed,->,line width=.9pt] (V2) -- ($(s2)!0.55!(V2)$);
\draw[gray,dashed,->,line width=.9pt] (W2) -- ($(U2)!0.55!(W2)$);

\node at (3,1) {\footnotesize \textbf{b)} Predecessors $\pi$: $\pi(U)=s$, $\pi(V)=s$, $\pi(W)=U$};
\end{scope}

\begin{scope}[yshift=-8cm]
\node[regular polygon,regular polygon sides=3,draw=black,line width=1pt,
      minimum size=16pt,inner sep=1pt,fill=yellow!25] (s3) at (0,3) {\small $s$};
\node[node] (U3) at (2,4) {\small $U$};
\node[node] (V3) at (2,2) {\small $V$};
\node[node,fill=green!12] (W3) at (4,3) {\small $W$};
\node[node,fill=green!12] (r3) at (6,3) {\small $r_i$};

\draw[->,line width=0.7pt,gray] (s3) .. controls +(0.8,-0.2) and +(-0.8,-0.2) .. (V3);
\draw[->,line width=0.7pt,gray] (V3) .. controls +(0.8,-0.3) and +(-0.8,-0.3) .. (W3);

\draw[->,blue,line width=1.6pt] (s3) .. controls +(0.8,0.2) and +(-0.8,0.2) .. (U3);
\draw[->,blue,line width=1.6pt] (U3) .. controls +(0.8,0.3) and +(-0.8,0.3) .. (W3);
\draw[->,blue,line width=1.6pt] (W3) .. controls +(0.8,0.2) and +(-0.8,0.2) .. (r3);

\node at (3,1) {\footnotesize \textbf{c)} Reconstructed shortest augmenting path $s\!\to\!U\!\to\!W\!\to\!r_i$};
\end{scope}

\end{tikzpicture}

%% file: figure_ek.tex
\begin{tikzpicture}[scale=.9,>=Stealth,
  node/.style={circle,draw=black,line width=1pt,minimum size=16pt,inner sep=1pt}]

  \node[regular polygon,regular polygon sides=3,draw=black,line width=1pt,
        minimum size=16pt,inner sep=1pt,fill=yellow!25] (sa) at (0,3) {\small $s$};
  \node[node] (Ua) at (2,4) {\small $U$};
  \node[node] (Va) at (2,2) {\small $V$};
  \node[node] (Wa) at (4,3) {\small $W$};
  \node[node,fill=green!12] (ra) at (6,3) {\small $r_i$};

  \draw[->,line width=1pt] (sa) .. controls +(0.8,0.20) and +(-0.8,0.20) .. (Ua); 
  \draw[->,line width=1pt] (Ua) .. controls +(-0.8,-0.20) and +(0.8,-0.20) .. (sa); 

  \draw[->,line width=1pt] (sa) .. controls +(0.8,-0.20) and +(-0.8,-0.20) .. (Va); 
  \draw[->,line width=1pt] (Va) .. controls +(-0.8,0.20) and +(0.8,0.20) .. (sa); 

  \draw[->,line width=1pt] (Ua) .. controls +(0.8,0.30) and +(-0.8,0.30) .. (Wa);  
  \draw[->,line width=1pt] (Wa) .. controls +(-0.8,-0.30) and +(0.8,-0.30) .. (Ua); 

  \draw[->,line width=1pt] (Va) .. controls +(0.8,-0.30) and +(-0.8,-0.30) .. (Wa); 
  \draw[->,line width=1pt] (Wa) .. controls +(-0.8,0.30) and +(0.8,0.30) .. (Va);  

  \draw[->,line width=1pt] (Wa) .. controls +(0.8,0.20) and +(-0.8,0.20) .. (ra);  
  \draw[->,line width=1pt] (ra) .. controls +(-0.8,-0.20) and +(0.8,-0.20) .. (Wa); 

  \draw[->,blue,line width=1.6pt] (sa) .. controls +(0.8,0.20) and +(-0.8,0.20) .. (Ua);
  \draw[->,blue,line width=1.6pt] (Ua) .. controls +(0.8,0.30) and +(-0.8,0.30) .. (Wa);
  \draw[->,blue,line width=1.6pt] (Wa) .. controls +(0.8,0.20) and +(-0.8,0.20) .. (ra);

  \node at (3,1) {\footnotesize \textbf{a)} Initial residual graph and shortest $s{\to}r_i$ path (blue)};

  \begin{scope}[yshift=-4cm]
    \node[regular polygon,regular polygon sides=3,draw=black,line width=1pt,
          minimum size=16pt,inner sep=1pt,fill=yellow!25] (sb) at (0,3) {\small $s$};
    \node[node] (Ub) at (2,4) {\small $U$};
    \node[node] (Vb) at (2,2) {\small $V$};
    \node[node] (Wb) at (4,3) {\small $W$};
    \node[node,fill=green!12] (rb) at (6,3) {\small $r_i$};

    \draw[->,line width=1pt] (Ub) .. controls +(-0.8,-0.20) and +(0.8,-0.20) .. (sb);

    \draw[->,line width=1pt] (sb) .. controls +(0.8,-0.20) and +(-0.8,-0.20) .. (Vb);
    \draw[->,line width=1pt] (Vb) .. controls +(-0.8,0.20) and +(0.8,0.20) .. (sb);

    \draw[->,line width=1pt] (Wb) .. controls +(-0.8,-0.30) and +(0.8,-0.30) .. (Ub);

    \draw[->,line width=1pt] (Vb) .. controls +(0.8,-0.30) and +(-0.8,-0.30) .. (Wb);
    \draw[->,line width=1pt] (Wb) .. controls +(-0.8,0.30) and +(0.8,0.30) .. (Vb);

    \draw[->,line width=1pt] (rb) .. controls +(-0.8,-0.20) and +(0.8,-0.20) .. (Wb);

    \node[red] at ($(sb)!0.5!(Ub)+(0,0.30)$) {\large $\times$};   
    \node[red] at ($(Ub)!0.5!(Wb)+(0,0.40)$) {\large $\times$};   
    \node[red] at ($(Wb)!0.5!(rb)+(0,0.30)$) {\large $\times$};   

    \node at (3,1) {\footnotesize \textbf{b)} Residual after augmentation: used forward arcs removed};
  \end{scope}
\end{tikzpicture}

%% file: online_example.tex

\tikzset{
node/.style={circle,draw,line width=0.9pt,minimum size=14pt,inner sep=0pt},
lab/.style={font=\footnotesize},
edg/.style={line width=1.1pt,->,>=Stealth},
}

\begin{tikzpicture}[scale=0.96]

\node[regular polygon,regular polygon sides=3,draw, line width=0.9pt,
      minimum size=14pt,inner sep=0pt,fill=yellow!25] (sa) at (0,5.2) {\small $s$};
\node[node] (Xa) at (-3.2,4.6) {$X$};
\node[node] (Ua) at (-2,3.6) {$U$};
\node[node] (Va) at ( 2,3.6) {$V$};
\node[node] (Wa) at ( 0,2.2) {$W$};
\node[node,fill=green!12] (r1a) at ( 0,0.6) {$r_1$};
\node[node,fill=green!12] (r2a) at ( 2,0.6) {$r_2$};
\node[node,fill=green!12] (r3a) at (-2,0.6) {$r_3$};

\draw[edg,gray!50] (sa) -- (Ua);
\draw[edg,gray!50] (sa) -- (Va);
\draw[edg,gray!50,bend left=14] (sa) to (Ua); 
\draw[edg,gray!50] (sa) -- (Xa);
\draw[edg,gray!50] (Ua) -- (Wa);
\draw[edg,gray!50] (Va) -- (Wa);
\draw[edg,gray!50] (Wa) -- (r1a);
\draw[edg,gray!50] (Wa) -- (r2a);
\draw[edg,gray!50] (Wa) -- (r3a);
\draw[edg,gray!50] (Ua) .. controls (-2,1.9) .. (r2a);
\draw[edg,gray!50] (Xa) .. controls (-3.0,3.2) and (-2.8,1.8) .. (r3a);

\draw[edg,blue] (sa) -- (Ua);
\draw[edg,blue] (Ua) -- (Wa);
\draw[edg,blue] (Wa) -- (r1a);
\draw[edg,darkgreen] (sa) -- (Va);
\draw[edg,darkgreen] (Va) -- (r1a);

\node[lab] at (0,-0.4) {(a) First receiver $r_1$: two edge-disjoint paths.};

\draw[dashed,gray!50,line width=0.4pt] (-4.0,-0.8) -- (4.0,-0.8);

\begin{scope}[yshift=-6.6cm]
\node[regular polygon,regular polygon sides=3,draw, line width=0.9pt,
      minimum size=14pt,inner sep=0pt,fill=yellow!25] (sb) at (0,5.2) {\small $s$};
\node[node] (Xb) at (-3.2,4.6) {$X$};
\node[node] (Ub) at (-2,3.6) {$U$};
\node[node] (Vb) at ( 2,3.6) {$V$};
\node[node] (Wb) at ( 0,2.2) {$W$};
\node[node,fill=green!12] (r1b) at ( 0,0.6) {$r_1$};
\node[node,fill=green!12] (r2b) at ( 2,0.6) {$r_2$};
\node[node,fill=green!12] (r3b) at (-2,0.6) {$r_3$};

\draw[edg,gray!50] (sb) -- (Ub);
\draw[edg,gray!50,bend left=14] (sb) to (Ub);
\draw[edg,gray!50] (sb) -- (Vb);
\draw[edg,gray!50] (sb) -- (Xb);
\draw[edg,gray!50] (Ub) -- (Wb);
\draw[edg,gray!50] (Vb) -- (Wb);
\draw[edg,gray!50] (Wb) -- (r1b);
\draw[edg,gray!50] (Wb) -- (r2b);
\draw[edg,gray!50] (Wb) -- (r3b);
\draw[edg,gray!50] (Ub) .. controls (-2,1.9) .. (r2b);
\draw[edg,gray!50] (Xb) .. controls (-3.0,3.2) and (-2.8,1.8) .. (r3b);

\draw[edg,blue] (sb) -- (Ub);
\draw[edg,blue] (Ub) -- (Wb);
\draw[edg,blue] (Wb) -- (r1b);
\draw[edg,darkgreen] (sb) -- (Vb);
\draw[edg,darkgreen] (Vb) -- (r1b);

\draw[edg,red,bend left=14] (sb) to (Ub);
\draw[edg,red] (Ub) .. controls (-2,1.9) .. (r2b);
\draw[edg,darkgreen] (Vb) -- (Wb);
\draw[edg,darkgreen] (Wb) -- (r2b);

\node[lab] at (0,-0.4) {(b) Second receiver $r_2$: one disjoint path  and one single-color overlap.};
\end{scope}

\draw[dashed,gray!50,line width=0.4pt] (-4.0,-7.4) -- (4.0,-7.4);

\begin{scope}[yshift=-13.2cm]
\node[regular polygon,regular polygon sides=3,draw, line width=0.9pt,
      minimum size=14pt,inner sep=0pt,fill=yellow!25] (sc) at (0,5.2) {\small $s$};
\node[node] (Xc) at (-3.2,4.6) {$X$};
\node[node] (Uc) at (-2,3.6) {$U$};
\node[node] (Vc) at ( 2,3.6) {$V$};
\node[node] (Wc) at ( 0,2.2) {$W$};
\node[node,fill=green!12] (r1c) at ( 0,0.6) {$r_1$};
\node[node,fill=green!12] (r2c) at ( 2,0.6) {$r_2$};
\node[node,fill=green!12] (r3c) at (-2,0.6) {$r_3$};

\draw[edg,gray!50] (sc) -- (Uc);
\draw[edg,gray!50,bend left=14] (sc) to (Uc);
\draw[edg,gray!50] (sc) -- (Vc);
\draw[edg,gray!50] (sc) -- (Xc);
\draw[edg,gray!50] (Uc) -- (Wc);
\draw[edg,gray!50] (Vc) -- (Wc);
\draw[edg,gray!50] (Wc) -- (r1c);
\draw[edg,gray!50] (Wc) -- (r2c);
\draw[edg,gray!50] (Wc) -- (r3c);
\draw[edg,gray!50] (Uc) .. controls (-2,1.9) .. (r2c);
\draw[edg,gray!50] (Xc) .. controls (-3.0,3.2) and (-2.8,1.8) .. (r3c);

\draw[edg,blue] (sc) -- (Uc);
\draw[edg,blue] (Uc) -- (Wc);
\draw[edg,blue] (Wc) -- (r1c);
\draw[edg,darkgreen] (sc) -- (Vc);
\draw[edg,darkgreen] (Vc) -- (r1c);
\draw[edg,darkgreen] (Vc) -- (Wc);
\draw[edg,darkgreen] (Wc) -- (r2c);
\draw[edg,red,bend left=14] (sc) to (Uc);
\draw[edg,red] (Uc) .. controls (-2,1.9) .. (r2c);

\draw[edg,blue] (Wc) -- (r3c);
\draw[edg,orange] (sc) -- (Xc);
\draw[edg,orange] (Xc) .. controls (-3.0,3.2) and (-2.8,1.8) .. (r3c);

\node[lab] at (0,-0.4) {(c) Third receiver $r_3$: two paths: one redirected and one disjoint.};
\end{scope}

\end{tikzpicture}

%% file: restrictedvsunrestricted.tex
\hspace*{-0.9cm}%
\begin{tikzpicture}[scale=.9,>=Stealth,
  node/.style={circle,draw=black,line width=1pt,minimum size=16pt,inner sep=1pt}]

\node[regular polygon,regular polygon sides=3,draw=black,line width=1pt,minimum size=16pt,inner sep=1pt,fill=yellow!25] (s0) at (0,3) {\small $s$};
\node[node] (a0) at (2,4) {};   
\node[node] (c0) at (2,2) {};   
\node[node] (b0) at (4,3) {};   
\node[node] (d0) at (5.4,3) {}; 
\node[node,fill=green!12,draw=black] (r10) at (7.2,4.2) {\small $r_1$};
\node[node,fill=green!12,draw=black] (r20) at (7.2,3.0) {\small $r_2$};
\node[node,fill=green!12,draw=black] (r30) at (7.2,1.8) {\small $r_3$};

\draw[->,line width=1pt,gray!55] (s0) -- (a0);
\draw[->,line width=1pt,gray!55] (s0) -- (c0);
\draw[->,line width=1pt,gray!55] (a0) -- (b0);
\draw[->,line width=1pt,gray!55] (c0) -- (b0);
\draw[->,line width=1pt,gray!55] (b0) -- (d0);
\draw[->,line width=1pt,gray!55] (a0) .. controls +(1.0,0.50) and +(-0.6,0.30) .. (r10);
\draw[->,line width=1pt,gray!55] (a0) .. controls +(1.0,-0.20) and +(-0.8,0.20) .. (r20);
\draw[->,line width=1pt,gray!55] (c0) .. controls +(1.0,0.20) and +(-0.8,-0.20) .. (r20);
\draw[->,line width=1pt,gray!55] (c0) .. controls +(1.0,-0.50) and +(-0.6,-0.30) .. (r30);
\draw[->,line width=1pt,gray!55] (d0) .. controls +(0.8,0.20) and +(-0.6,-0.10) .. (r10);
\draw[->,line width=1pt,gray!55] (d0) .. controls +(0.8,-0.20) and +(-0.6,0.10) .. (r30);

\node at (3.6,0.7) {\footnotesize \textbf{a)} Topology before algorithms.};


\begin{scope}[yshift=-5cm]
\node[regular polygon,regular polygon sides=3,draw=black,line width=1pt,minimum size=16pt,inner sep=1pt,fill=yellow!25] (sa) at (0,3) {\small $s$};
\node[node] (aa) at (2,4) {}; \node[node] (ca) at (2,2) {}; \node[node] (ba) at (4,3) {}; \node[node] (da) at (5.4,3) {};
\node[node,fill=green!12,draw=black] (r1a) at (7.2,4.2) {\small $r_1$};
\node[node,fill=green!12,draw=black] (r2a) at (7.2,3.0) {\small $r_2$};
\node[node,fill=green!12,draw=black] (r3a) at (7.2,1.8) {\small $r_3$};

\draw[->,line width=1pt,gray!55] (sa) -- (aa);
\draw[->,line width=1pt,gray!55] (sa) -- (ca);
\draw[->,line width=1pt,gray!55] (aa) -- (ba);
\draw[->,line width=1pt,gray!55] (ca) -- (ba);
\draw[->,line width=1pt,gray!55] (ba) -- (da);
\draw[->,line width=1pt,gray!55] (aa) .. controls +(1.0,0.50) and +(-0.6,0.30) .. (r1a);
\draw[->,line width=1pt,gray!55] (aa) .. controls +(1.0,-0.20) and +(-0.8,0.20) .. (r2a);
\draw[->,line width=1pt,gray!55] (ca) .. controls +(1.0,0.20) and +(-0.8,-0.20) .. (r2a);
\draw[->,line width=1pt,gray!55] (ca) .. controls +(1.0,-0.50) and +(-0.6,-0.30) .. (r3a);
\draw[->,line width=1pt,gray!55] (da) .. controls +(0.8,0.20) and +(-0.6,-0.10) .. (r1a);
\draw[->,line width=1pt,gray!55] (da) .. controls +(0.8,-0.20) and +(-0.6,0.10) .. (r3a);

\draw[->,blue,line width=1.6pt] (sa) .. controls +(0.9,0.60) and +(-0.6,0.55) .. (aa);
\draw[->,blue,line width=1.6pt] (aa) .. controls +(1.0,0.50) and +(-0.6,0.30) .. (r1a);
\draw[->,blue,densely dashed,line width=1.6pt] (sa) .. controls +(0.9,-0.40) and +(-0.6,-0.35) .. (ca);
\draw[->,blue,densely dashed,line width=1.6pt] (ca) -- (ba);
\draw[->,blue,densely dashed,line width=1.6pt] (ba) .. controls +(0.9,0.18) and +(-0.7,0.18) .. (da);
\draw[->,blue,densely dashed,line width=1.6pt] (da) .. controls +(0.8,0.20) and +(-0.6,-0.10) .. (r1a);

\draw[->,green!50!black,line width=1.6pt] (sa) .. controls +(0.9,0.35) and +(-0.6,0.30) .. (aa);
\draw[->,green!50!black,line width=1.6pt] (aa) .. controls +(1.0,-0.20) and +(-0.8,0.20) .. (r2a);
\draw[->,green!50!black,densely dashed,line width=1.6pt] (sa) .. controls +(0.9,-0.15) and +(-0.6,-0.10) .. (ca);
\draw[->,green!50!black,densely dashed,line width=1.6pt] (ca) .. controls +(1.0,0.20) and +(-0.8,-0.20) .. (r2a);

\draw[->,orange,line width=1.6pt] (sa) .. controls +(0.9,-0.65) and +(-0.6,-0.55) .. (ca);
\draw[->,orange,line width=1.6pt] (ca) .. controls +(1.0,-0.50) and +(-0.6,-0.30) .. (r3a);
\draw[->,orange,densely dashed,line width=1.6pt] (sa) .. controls +(0.9,0.10) and +(-0.6,0.05) .. (aa);
\draw[->,orange,densely dashed,line width=1.6pt] (aa) -- (ba);
\draw[->,orange,densely dashed,line width=1.6pt] (ba) .. controls +(0.9,-0.18) and +(-0.7,-0.18) .. (da);
\draw[->,orange,densely dashed,line width=1.6pt] (da) .. controls +(0.8,-0.20) and +(-0.6,0.10) .. (r3a);

\node at (3.6,0.7) {\footnotesize \textbf{b)} Per-receiver EK: each $r_i$ attains two edge-disjoint paths.};
\end{scope}


\begin{scope}[yshift=-10cm]
\node[regular polygon,regular polygon sides=3,draw=black,line width=1pt,minimum size=16pt,inner sep=1pt,fill=yellow!25] (sb) at (0,3) {\small $s$};
\node[node] (ab) at (2,4) {}; \node[node] (cb) at (2,2) {}; \node[node] (bb) at (4,3) {}; \node[node] (db) at (5.4,3) {};
\node[node,fill=green!12,draw=black] (r1b) at (7.2,4.2) {\small $r_1$};
\node[node,fill=green!12,draw=black] (r2b) at (7.2,3.0) {\small $r_2$};
\node[node,fill=green!12,draw=black] (r3b) at (7.2,1.8) {\small $r_3$};

\draw[->,line width=1pt,gray!55] (sb) -- (ab);
\draw[->,line width=1pt,gray!55] (sb) -- (cb);
\draw[->,line width=1pt,gray!55] (ab) -- (bb);
\draw[->,line width=1pt,gray!55] (cb) -- (bb);
\draw[->,line width=1pt,gray!55] (bb) -- (db);
\draw[->,line width=1pt,gray!55] (ab) .. controls +(1.0,0.50) and +(-0.6,0.30) .. (r1b);
\draw[->,line width=1pt,gray!55] (ab) .. controls +(1.0,-0.20) and +(-0.8,0.20) .. (r2b);
\draw[->,line width=1pt,gray!55] (cb) .. controls +(1.0,0.20) and +(-0.8,-0.20) .. (r2b);
\draw[->,line width=1pt,gray!55] (cb) .. controls +(1.0,-0.50) and +(-0.6,-0.30) .. (r3b);
\draw[->,line width=1pt,gray!55] (db) .. controls +(0.8,0.20) and +(-0.6,-0.10) .. (r1b);
\draw[->,line width=1pt,gray!55] (db) .. controls +(0.8,-0.20) and +(-0.6,0.10) .. (r3b);

\draw[->,blue,line width=1.8pt] (sb) -- (ab);
\draw[->,blue,line width=1.8pt] (ab) .. controls +(1.0,0.50) and +(-0.6,0.30) .. (r1b);
\draw[->,blue,line width=1.8pt] (ab) .. controls +(1.0,-0.20) and +(-0.8,0.20) .. (r2b);

\draw[->,orange,line width=1.8pt] (sb) -- (cb);
\draw[->,orange,line width=1.8pt] (cb) -- (bb);
\draw[->,orange,line width=1.8pt] (bb) -- (db);
\draw[->,orange,line width=1.8pt] (cb) .. controls +(1.0,0.20) and +(-0.8,-0.20) .. (r2b);
\draw[->,orange,line width=1.8pt] (db) .. controls +(0.8,0.20) and +(-0.6,-0.10) .. (r1b);
\draw[->,orange,line width=1.8pt] (db) .. controls +(0.8,-0.20) and +(-0.6,0.10) .. (r3b);

\node at (3.6,0.7) {\footnotesize \textbf{c)} Restricted EK (zones): $r_1,r_2$ use blue and orange; $r_3$ only orange.};
\end{scope}

\end{tikzpicture}

%% file: performance_1_WS.tex
\begin{tikzpicture}
  \begin{axis}[
    width=\columnwidth, height=5.2cm,
    xlabel={Nodes $n$}, ylabel={Group multicast max-flow},
    grid=both, grid style={gray!20},
    legend pos=north west, legend cell align={left},
    label style={font=\small}, tick label style={font=\small},
    every axis plot/.append style={line width=1pt, mark size=1.6pt},
    cycle list={{solid,mark=*},{densely dashed,mark=square*},{dotted,mark=triangle*}},
    mark repeat=6,
  ]


    \addplot table[
      x index=0, y index=3, col sep=space, comment chars=\#
    ] {ejecucion_online_beta_0.10_densidad_0.10_densidad_recep_0.25_zonas.dat};
    \addlegendentry{Edge density 10\%}

    \addplot table[
      x index=0, y index=3, col sep=space, comment chars=\#
    ] {ejecucion_online_beta_0.10_densidad_0.20_densidad_recep_0.25_zonas.dat};
    \addlegendentry{Edge density 20\%}

    \addplot table[
      x index=0, y index=3, col sep=space, comment chars=\#
    ] {ejecucion_online_beta_0.10_densidad_0.30_densidad_recep_0.25_zonas.dat};
    \addlegendentry{Edge density 30\%}


  \end{axis}
\end{tikzpicture}
\caption{WS ($\beta=0.1$), receiver density $25\%$. 
Group multicast max-flow vs.\ $n$ for \textbf{edge densities} $10\%$, $20\%$, and $30\%$. Restricted EK matches EK (curves overlap).}

%% file: performance_1_ER.tex
\begin{tikzpicture}
  \begin{axis}[
    width=\columnwidth, height=5.2cm,
    xlabel={Nodes $n$}, ylabel={Group multicast max-flow},
    grid=both, grid style={gray!20},
    legend pos=north west, legend cell align={left},
    label style={font=\small}, tick label style={font=\small},
    every axis plot/.append style={line width=1pt, mark size=1.6pt},
    cycle list={{solid,mark=*},{densely dashed,mark=square*},{dotted,mark=triangle*}},
    mark repeat=6,
  ]


    \addplot table[
      x index=0, y index=3, col sep=space, comment chars=\#
    ] {ejecucion_online_renyi_densidad_0.10_densidad_recep_0.25_zonas.dat};
    \addlegendentry{Edge density 10\%}

    \addplot table[
      x index=0, y index=3, col sep=space, comment chars=\#
    ] {ejecucion_online_renyi_densidad_0.20_densidad_recep_0.25_zonas.dat};
    \addlegendentry{Edge density 20\%}

    \addplot table[
      x index=0, y index=3, col sep=space, comment chars=\#
    ] {ejecucion_online_renyi_densidad_0.30_densidad_recep_0.25_zonas.dat};
    \addlegendentry{Edge density 30\%}


  \end{axis}
\end{tikzpicture}
\caption{ER, receiver density $25\%$. 
Group multicast max-flow vs.\ $n$ for \textbf{edge densities} $10\%$, $20\%$, and $30\%$. Online matches EK (curves overlap).}

%% file: performance_2.tex

\begin{tikzpicture}
  \begin{axis}[
    width=\columnwidth, height=5.2cm,
    xlabel={Nodes $n$}, ylabel={Group multicast max-flow},
    grid=both, grid style={gray!20},
    legend pos=north west, legend cell align={left},
    label style={font=\small}, tick label style={font=\small},
    every axis plot/.append style={line width=1pt, mark size=1.8pt},
    cycle list={{solid,mark=*},{densely dashed,mark=square*}},
    mark repeat=1,                    
    restrict x to domain=800:900,     
    unbounded coords=discard,         
    ymin=1.7, ymax=4.3, ytick={2,3,4},  
    xtick distance=10,
  ]

    \addplot table[
      x index=0, y index=3, col sep=space, comment chars=\#
    ] {ejecucion_4k_online_beta_0.10_densidad_0.40_densidad_recep_0.30_zonas.dat};
    \addlegendentry{EK (unconstrained)}

    \addplot table[
      x index=0, y index=4, col sep=space, comment chars=\#
    ] {ejecucion_4k_online_beta_0.10_densidad_0.40_densidad_recep_0.30_zonas.dat};
    \addlegendentry{Restricted EK}

    \addplot[only marks,mark=*,mark size=2pt,black,forget plot]
      table[x index=0,
            y expr=(\thisrowno{3}-\thisrowno{4})>0 ? \thisrowno{4} : nan,
            col sep=space]{ejecucion_4k_online_beta_0.10_densidad_0.40_densidad_recep_0.30_zonas.dat};
    \addplot[only marks,mark=*,mark size=2pt,black,forget plot]
      table[x index=0,
            y expr=(\thisrowno{3}-\thisrowno{4})>0 ? \thisrowno{3} : nan,
            col sep=space]{ejecucion_4k_online_beta_0.10_densidad_0.40_densidad_recep_0.30_zonas.dat};


  \end{axis}
\end{tikzpicture}

\caption{WS (\(k{=}4\), \(\beta{=}0.1\)). Group multicast max-flow vs.\ \(n\), \emph{zoom} a \(n\in[800,900]\).
The $n=840$ case shows a difference between both algorithms.}

%% file: runtime_ratio.tex
\begin{tikzpicture}
  \begin{axis}[
    width=\columnwidth, height=5.2cm,
    xlabel={Nodes $n$}, ylabel={Runtime Ratio (Online vs EK)},
    grid=both, grid style={gray!20},
    legend pos=north east, legend cell align={left},
    label style={font=\small}, tick label style={font=\small},
    every axis plot/.append style={line width=1pt, mark size=1.6pt},
    cycle list={{solid,mark=*},{densely dashed,mark=square*},{dotted,mark=triangle*}},
    mark repeat=6,
    ymin=0, ymax=2, 
    legend style={
      font=\scriptsize,
      fill=white,
      fill opacity=0.9,
      draw=black,
      rounded corners=2pt
    }
  ]


    \addplot[black, dashed, line width=0.8pt, domain=0:200, samples=2] {1};
    \addlegendentry{Parity ($=1$)}

    \addplot table[
      x index=0, y index=5, col sep=space, comment chars=\#
    ] {ejecucion_time_online_beta_0.10_densidad_0.10_densidad_recep_0.25_zonas.dat};
    \addlegendentry{Edge density 10\%}

    \addplot table[
      x index=0, y index=5, col sep=space, comment chars=\#
    ] {ejecucion_time_online_beta_0.10_densidad_0.20_densidad_recep_0.25_zonas.dat};
    \addlegendentry{Edge density 20\%}

    \addplot table[
      x index=0, y index=5, col sep=space, comment chars=\#
    ] {ejecucion_time_online_beta_0.10_densidad_0.30_densidad_recep_0.25_zonas.dat};
    \addlegendentry{Edge density 30\%}


  \end{axis}
\end{tikzpicture}
\caption{Runtime ratio of the online construction relative to the per-receiver Edmonds–Karp baseline for Watts–Strogatz graphs ($\beta=0.1$) with a receiver density of 25\%. The curves correspond to edge densities of 10\%, 20\%, and 30\%. The dashed line indicates parity ($=1$); values below one denote faster execution of the online algorithm.}
\label{fig:ws_025}

%% file: SourceCoding.bbl
\begin{thebibliography}{10}
\providecommand{\url}[1]{#1}
\csname url@samestyle\endcsname
\providecommand{\newblock}{\relax}
\providecommand{\bibinfo}[2]{#2}
\providecommand{\BIBentrySTDinterwordspacing}{\spaceskip=0pt\relax}
\providecommand{\BIBentryALTinterwordstretchfactor}{4}
\providecommand{\BIBentryALTinterwordspacing}{\spaceskip=\fontdimen2\font plus
\BIBentryALTinterwordstretchfactor\fontdimen3\font minus \fontdimen4\font\relax}
\providecommand{\BIBforeignlanguage}[2]{{%
\expandafter\ifx\csname l@#1\endcsname\relax
\typeout{** WARNING: IEEEtran.bst: No hyphenation pattern has been}%
\typeout{** loaded for the language `#1'. Using the pattern for}%
\typeout{** the default language instead.}%
\else
\language=\csname l@#1\endcsname
\fi
#2}}
\providecommand{\BIBdecl}{\relax}
\BIBdecl

\bibitem{moy1994multicast}
J.~Moy, ``Multicast routing in datagram internetworks and extended lans,'' \emph{ACM SIGCOMM Computer Communication Review}, vol.~24, no.~4, pp. 55--64, 1994.

\bibitem{widmer2001extending}
J.~Widmer and J.~L. Boudec, ``Extending a multicast tree using diversity-based routing,'' in \emph{Proceedings IEEE INFOCOM 2001}, vol.~2.\hskip 1em plus 0.5em minus 0.4em\relax IEEE, 2001, pp. 924--933.

\bibitem{chu2002case}
Y.~Chu, S.~G.~Rao, and H.~Zhang, ``A case for end system multicast,'' in \emph{ACM SIGMETRICS Performance Evaluation Review}, vol.~30, no.~1.\hskip 1em plus 0.5em minus 0.4em\relax ACM, 2002, pp. 1--12.

\bibitem{Ford1956}
L.~R. Ford and D.~R. Fulkerson, ``Maximal flow through a network,'' \emph{Canadian J. of Mathematics}, vol.~8, pp. 399--404, 1956.

\bibitem{Dinic1970}
E.~A. Dinic, ``Algorithm for solution of a problem of maximum flow in networks with power estimation,'' \emph{Soviet Mathematics Doklady}, vol.~11, pp. 1277--1280, 1970.

\bibitem{EdmondsKarp1972}
J.~Edmonds and R.~M. Karp, ``Theoretical improvements in algorithmic efficiency for network flow problems,'' \emph{Journal of the ACM}, vol.~19, no.~2, pp. 248--264, 1972.

\bibitem{ahlswede2000network}
R.~Ahlswede, N.~Cai, S.-Y.~R. Li, and R.~W. Yeung, ``Network information flow,'' \emph{IEEE Transactions on Information Theory}, vol.~46, no.~4, pp. 1204--1216, 2000.

\bibitem{Li2003}
S.-Y. Li, R.~Yeung, and N.~Cai, ``Linear network coding,'' \emph{{IEEE} Trans. on Inf. Theory}, vol.~49, no.~2, pp. 371--381, feb 2003.

\bibitem{Koetter2003}
R.~Koetter and M.~Medard, ``An algebraic approach to network coding,'' \emph{{IEEE}/{ACM} Trans. on Networking}, vol.~11, no.~5, pp. 782--795, oct 2003.

\bibitem{chou2003practical}
P.~A. Chou, Y.-H. Wu, and K.~Jain, ``Practical network coding,'' in \emph{Annual Allerton Conference on Communication, Control, and Computing}, 2003, pp. 40--49.

\bibitem{ho2006random}
T.~Ho, M.~Medard, R.~Koetter, D.~R. Karger, M.~Effros, J.~Shi, and B.~Leong, ``A random linear network coding approach to multicast,'' in \emph{IEEE Transactions on Information Theory}, vol.~52, no.~10.\hskip 1em plus 0.5em minus 0.4em\relax IEEE, 2006, pp. 4413--4430.

\bibitem{lestayo2001adaptive}
T.~Lestayo, V.~M. Fern'andez, and C.~L\'opez, ``Adaptive approach for fec reliable multicast,'' \emph{Electronics Letters}, vol.~37, no.~22, pp. 1333--1335, Oct. 2001.

\bibitem{Lestayo2023}
T.~Lestayo-Martínez and M.~Fernández-Veiga, ``Source-coded multicast for efficient content delivery,'' \emph{2023 IEEE 28th International Workshop on Computer Aided Modeling and Design of Communication Links and Networks (CAMAD)}, pp. 140--145, 2023.

\bibitem{lestayo2024source}
T.~Lestayo and M.~Fern{\'a}ndez, ``Source-coded multicast with single and aggregated sources for efficient content delivery,'' \emph{IEEE Open Journal of the Communications Society}, 2024.

\bibitem{steiner1991approximation}
D.-Z. Du and F.~K. Hwang, ``Approximation algorithms for the steiner tree problem in networks,'' in \emph{Networks}, vol.~21.\hskip 1em plus 0.5em minus 0.4em\relax Wiley Online Library, 1991, pp. 73--82.

\bibitem{Hwang1992}
F.~K. Hwang and D.~S. Richards, ``Steiner tree problems,'' \emph{Networks}, vol.~22, no.~1, pp. 55--89, Jan. 1992.

\bibitem{Charikar1999}
M.~Charikar, C.~Chekuri, T.-y. Cheung, Z.~Dai, A.~Goel, S.~Guha, and M.~Li, ``Approximation algorithms for directed steiner problems,'' \emph{Journal of Algorithms}, vol.~33, no.~1, pp. 73--91, Oct. 1999.

\bibitem{rouskas1997multicast}
G.~N. Rouskas, ``Multicast routing and wavelength assignment in optical networks,'' in \emph{Proceedings of IEEE INFOCOM}.\hskip 1em plus 0.5em minus 0.4em\relax IEEE, 1997, pp. 188--195.

\bibitem{Ahlswede2000}
R.~Ahlswede, N.~Cai, S.-Y. Li, and R.~Yeung, ``Network information flow,'' \emph{{IEEE} Trans. on Inf. Theory}, vol.~46, no.~4, pp. 1204--1216, jul 2000.

\bibitem{lun2005network}
D.~Lun, N.~Ratnakar, R.~Koetter, and M.~Médard, ``Network coding for efficient wireless unicast,'' in \emph{Proceedings of the 43rd Annual Allerton Conference on Communication, Control, and Computing}, 2005, pp. 621--630.

\bibitem{padmanabhan2003resilient}
V.~N. Padmanabhan, L.~Qiu, and G.~M. Voelker, ``Resilient peer-to-peer streaming,'' in \emph{Proceedings of the 2nd International Workshop on Peer-to-Peer Systems (IPTPS)}, 2003, pp. 73--82.

\bibitem{kandula2005walking}
S.~Kandula, D.~Katabi, B.~Davie, and H.~Chiu, ``Walking the tightrope: responsive yet stable traffic engineering,'' in \emph{ACM SIGCOMM 2005}.\hskip 1em plus 0.5em minus 0.4em\relax ACM, 2005, pp. 253--264.

\bibitem{Erdos1959}
P.~Erdős and A.~Rényi, ``On random graphs i,'' \emph{Publicationes Mathematicae}, vol.~6, pp. 290--297, 1959.

\bibitem{Watts1998}
D.~J. Watts and S.~H. Strogatz, ``Collective dynamics of `small-world' networks,'' \emph{Nature}, vol. 393, pp. 440--442, 1998.

\end{thebibliography}
